\documentclass{aa}
\usepackage{graphicx}
\usepackage{rotating}
\usepackage{lscape}
\usepackage{txfonts}
\begin{document}
   \title{The intermediate-redshift galaxy cluster CL~0048-2942.}

   \subtitle{Stellar populations.}

   \author{M. Serote Roos\inst{1}, C. Lobo\inst{2,3},
F. Durret\inst{4}, A. Iovino\inst{5} \and
I. M\'arquez\inst{6}
          }

   \offprints{M. Serote Roos}

   \institute{Centro de Astronomia e Astrof\'{\i}sica da Universidade de 
Lisboa, Tapada da Ajuda, P-1349-018 Lisboa, Portugal \\
\email{serote@oal.ul.pt}
         \and
Depto. de Matem\'atica Aplicada, Faculdade de Ci\^encias, Univ. do Porto, R. do Campo Alegre 687, P-4169-007 Porto, Portugal
\and Centro de Astrof\'{\i}sica da Universidade do Porto, 
Rua das Estrelas, P-4150-762 Porto, Portugal\\
\email{lobo@astro.up.pt}
\and
Institut d'Astrophysique de Paris, 98bis Bd. Arago, F-75014 Paris, France\\
\email{durret@iap.fr} 
\and  Osservatorio Astronomico di Brera, via Brera 28, 20121 Milano, Italy\\
\email{iovino@brera.mi.astro.it}
\and Instituto de Astrof\'{\i}sica de Andaluc\'{\i}a (CSIC), Camino Bajo
de Hu\'etor, 24, 18008 Granada, Spain\\
\email{isabel@iaa.es}
            }

   \date{Received ; }

\abstract{We present a detailed study of the cluster CL~0048-2942,
located at $z \sim 0.64$, based on a photometric and spectroscopic
catalogue of $54$ galaxies in a $5 \times 5$ arcmin$^2$ region centred
in that cluster. Of these, 23 galaxies were found to belong to the 
cluster. Based on this sample, 
the line-of-sight velocity dispersion of the cluster is approximately $680$
$\pm$ 140 km/s.
 We have performed stellar population synthesis in the
cluster members as well as in the field galaxies of the sample and
found that there are population gradients in the cluster with central
galaxies hosting mainly intermediate/old populations whereas galaxies
in the cluster outskirts show clearly an increase of younger
populations, meaning that star formation is predominantly taking place
in the outer regions of the cluster.  In a general way, field galaxies
seem to host less evolved stellar populations than cluster members.
In fact, in terms of ages, young supergiant stars dominate the spectra
of field galaxies whereas cluster galaxies display a dominant number
of old and intermediate age stars.  Following the work of other
authors (e.g. Dressler et al. 1999) we have estimated the percentage
of K+A galaxies in our sample
and found around
13\% in the cluster and 10\% in the field. These values were estimated
through means of a new  method, based on stellar population synthesis results,
that takes into account all possible absorption features in the spectrum
and thus makes optimal use of the data.


   \keywords{galaxies: clusters: CL~0048-2942 -- 
galaxies: clusters: general -- galaxies: distances and redshifts ---
galaxies: stellar content
               }
   }
  \titlerunning{Stellar populations in CL~0048-2942}
  \authorrunning{Serote Roos et al.}

   \maketitle
%

\section{Introduction}

It has now become a well-accepted fact that
the surrounding environment strongly affects the
properties of galaxies, in particular their stellar content. Cluster
members will thus reveal different stellar populations when compared
to field galaxies. One could try to explain this discrepancy by simply
invoking the different morphological mix of the two environments: rich
clusters or their denser inner regions are, in general, dominated by
lenticulars and ellipticals whereas the field is more abundant in
spirals (Oemler 1974). This is actually the so-called morphology-density relation (Dressler 1980 and see Smith et al.
2004 and references therein). Whether the morphological mix itself is
driven by environment (``nurture'' scenario) or set by the initial
density conditions of the region where the galaxies are born
(``nature'' scenario) is another issue. General consensus has been
reached in accepting the two factors as relevant ones, and now the
major debate has shifted into determining the relative importance of
each one of them and pinning down which environmental mechanisms
contribute significantly to morphological evolution and where (e.g.
Goto et al. 2004, Smith et al. 2004 and references mentioned in these
works).

Moreover, a quite natural expectation is that, even within a same
morphological class (or a proxy for it),
different star formation histories will be experienced by galaxies
inhabiting the two environments. This has been observed (see e.g.
Ellingson et al. 2001) and happens even when cluster sampling comes
from less evolved or less dense clusters: a flagrant example is
provided by the anaemic Virgo spirals (e.g. Kennicutt, 1983).

So, both parameters - morphology and star formation history, which is
betrayed by the galaxy spectrum - must be taken into account 
to provide a reasonable and complete answer on how (physically and
quantitatively) environment affects the stellar contents of galaxies.

A probable evidence of environmentally-driven evolution could be
provided by the non negligible fraction of post-star forming galaxies
(i.e. the strong Balmer absorption galaxies, also named post-starburst,
K+A, k+a/a+k or even E+A galaxies; Dressler \& Gunn 1983, Couch \&
Sharples 1987, Balogh et al. 1999, Tran et al.  2003 and references
therein, Goto et al. 2003) that clusters at intermediate redshifts
have been observed to host. And even if the relative percentage of
these galaxies in the field seems to be lower than in clusters (see
Tran et al. 2004 and references therein), establishing what this class
represents in terms of spectral evolution of galaxies in different
environments is a delicate task, far from being settled, and still in
need of more data to allow one to draw conclusive inferences.

Last but not least, and apart from the evident distinction patent
between clusters and field, the observations and results on stellar
populations of cluster galaxies are certainly dependent on several
factors that can be either physical or the effect of biases. Among
such parameters one can enumerate the mass or richness of the cluster,
its dynamical state and evolutionary state of the intra-cluster
medium, the extent of the physical region we are probing within the
cluster, the accretion rate of infalling galaxies entering the
cluster, and even, when comparing different systems, 
the detection method that was used to select a cluster.

In order to disentangle real effects from biases and establish the
dependencies of stellar populations on physical parameters, more
systems covering a range of the above mentioned properties need to be
thoroughly studied.

In this paper we present an analysis of the galaxy cluster
CL~0048-2942, located at redshift $z \sim 0.64$. Based on photometric
and spectroscopic data, we have analysed some of the cluster
properties (general shape, velocity dispersion, evidence for
substructure), as well as of its member galaxies (stellar content,
colours and spectral properties) as a function of the radial
distribution inside the cluster.

This system falls in a redshift range that has been sparsely sampled
and studied up to now.  In fact, and though allowing one to obtain
ample spatial coverage for numerous clusters and adjacent fields,
cluster spectroscopic surveys do not usually allow us to reach systems
located as far away as our cluster (e.g. the CNOC1 survey of Yee,
Ellingson \& Carlberg 1996, probing $0.18 \la$ z $\la 0.55$).
Alternatively, spectroscopic studies have been carried out in massive
notorious individual systems that were pin-pointed at high (z $>0.8$)
redshifts. These were selected from X-ray surveys (e.g.  X-ray bright
sources MS$1054-0321$ at z $\simeq 0.83$ unveiled in the EMSS by
Luppino \& Gioia 1995, Donahue et al. 1998; and RX J$0848.9+4452$
located at z $=1.24$ and detected by Rosati et al. 1999 with ROSAT; Cl
J$0152.7-1357$ at z $= 0.835$ and Cl J$1226.9+3332$ at z $\simeq 0.89$
found in the WARPS survey by Della Ceca et al. 2000, Ebeling et al.
2000, Ebeling et al.  2001); around radio galaxies acting as signposts
for high density regions (e.g.  Deltorn et al. 1997 uncovered a
z $=0.996$ cluster around $3$CR $184$; Blanton et al. 2003 probed the
surroundings of a VLA FIRST radio source identifying a cluster at
z $=0.96$); or by targeting high S/N candidates selected in deep
photometric surveys (as is the case of clusters $0023+0423$ and
$1604+4304$ at redshifts $0.84$ and $0.90$, respectively, that are
being studied by the PDCS group - Postman et al. 1998). Obvious
observational limitations imply that detailed studies like the ones
performed at low-z remain, up to now, the exception to the rule when
one refers to these intermediate and distant clusters (e.g. see the
work of van Dokkum et al. 1998, 1999 on cluster MS$1054-0321$ and the
follow-up of a handful of high-redshift clusters by the PDCS team;
Postman et al. 1998, Lubin et al. 1998, Postman et al. 2001, Lubin et
al. 2002).

These and several other similar examples suggest a redshift slice
between $z \simeq 0.5$ and $z \simeq 0.8$ still largely unexplored in
what concerns cluster studies. And this paucity actually spreads,
apart from some sparse examples such as the ones mentioned above, up
to $z \sim 1 - 1.3$ (the approximate limit of cluster detections up
to now).  This scarcity is even more important since (1) the higher
half of this interval roughly corresponds to an epoch of rapid cluster
building, mainly through the infall of groups according to
hierarchical theories of structure formation (Kauffmann 1995,
corrected for the actually most favoured $\Lambda$-dominated Universe)
or of individual disk field galaxies (Dressler 2004); (2) the lower
limits of it regard the period where clusters are undergoing (or have
recently completed) virialisation (Postman et al.  1998); (3) $z \sim
1$ is the approximate limit of the epoch where star-forming activity
across the Universe starts declining, as determined by studies of the
global star formation rate (Lilly et al.  1996, Madau et al. 1996,
1998, Hartwick 2004 and references therein).

The study of varied types of clusters at these redshifts is thus,
apart from fascinating, an urging need (and guarantee) for getting
closer to unravelling the star formation histories of galaxies as well
as the evolution of structures (from galaxies to clusters).

In section~\ref{sec_obs} we describe the photometric and spectroscopic
observations of the cluster field and respective data reduction. In
section~\ref{sec_zcat} we give a brief account on the redshift
measures obtained for the galaxies in the field-of-view of
CL~0048-2942 and present the final catalogue for that field. The
redshift interval that defines the cluster is determined in
section~\ref{sec_clus}, where we also discuss some of its structural
and dynamical properties.  Section~\ref{sec_starpop} deals with the
stellar population synthesis, describing the method used, its
application to the sample and the results obtained.  Spectral
classifications are made in section~\ref{sec_ka} and results are
discussed and compared to the stellar synthesis ones. In
section~\ref{sec_conc} we discuss these results, together with some
colour analysis, and draw some conclusions, while in
section~\ref{sec_sum} we summarize this work.

\section{Observations and data reduction}\label{sec_obs}

CL~0048-2942 was identified in optical images of the ESO Imaging
Survey (Nonino et al. 1999) by running a matched-filter detection
algorithm (Lobo et al. 2000) on the I-band of that data set. It was
actually one of the most promising members of the catalogue of cluster
candidates of Lobo et al. (2000) due to its high S/N of detection on
the I-band image, being one of the richest and most distant systems of
that catalogue: a very rough and tentative estimate of its redshift,
performed with the detection algorithm, initially placed it at $z \sim
0.85$ (matched-filter algorithms generally over-estimate cluster
redshifts - Ramella et al. 2000). Deeper multi-colour follow-up
photometry allowed us to confirm the reality of the cluster as an
excess number density of galaxies with coherent colours (Andreon et
al. 2004), and to select targets for follow-up spectroscopy.

\subsection{Photometry}\label{sec_phot}

Images of the cluster field (each one covering $\sim 5 \times 5$
square arcmin) were obtained through the B, V, R and I Bessel-Cousins
filters at the La Silla ESO $3.6$m telescope with the EFOSC2 camera.
Standard data reduction was performed with IRAF. Photometry was done
with {\em SExtractor} (Bertin \& Arnouts 1996) providing magnitudes
within a $4.5$ arcsec aperture that were calibrated to the Landolt
(1992) system, and colours for all galaxies detected in the I-band
images, with a completeness down to I $\sim 23$. All details on
photometry are provided in Andreon et al. (2004).

We limited the catalogue to $20.0 <$ I $<22.0$, having in mind the
typical range of magnitudes expected for interemediate redshifts
(ie. excluding bright objects). Then, and even though we expected a
low contamination rate by stars (CL~0048-2942 lies in a direction
close to the south galactic pole), we used a joint criterion based on
surface brightness (compactness) and shape (stellarity index) to order
targets in preferential order for spectroscopy. This order was however
not a strict constraint due to the limitations imposed by MOS mask
construction (see next section).


\subsection{Spectroscopy}\label{sec_spectro}

Multi-object spectroscopy was performed during two runs at the VLT: a
first campaign of two nights in September 1999 with the MOS unit of
FORS1 (VLT UT1) provided partial coverage of selected objects in the
cluster candidate field; further data were gathered during the second
campaign, this time in service mode during the first semester of 2000,
with the MOS unit of FORS2 (VLT UT2). We used the grism 300V+10 in the
first run and the practically identical grism 300V+20 in the second run,
both times applying also the order separation filter GG435. This setup
provided a useful field of view of $4.7 \times 6.8$ arcmin$^2$,
covering the spectral range $4450-8650$ \AA. The dispersion of $112$
\AA/mm ($2.69$ \AA/pixel) gave a spectral resolution of $500$ or about
$16$ \AA\ for a slit width of $1.2$ arcsec.

All nights were photometric, with mean seeing of $0.8''$ 
during both runs. For the first run typical airmass values were
$1.1 - 1.2$, whereas for the second run these values ranged between
$1.1$ and $2$.  Table~\ref{tab_obslog} reproduces the log of the
observations.

\begin{table*}
\caption{Log of the observations.}
\begin{center}
\begin{tabular}{rrrrrr}
\hline
Run/Instrument  &  Night  &  Seeing [arcsec]  &  Airmass  &  Photometric conditions & Observation mode \\
\hline     
1/FORS1 & 12-09-1999  & 0.9--1.2  & 1.00--1.20 & photometric & visitor \\
1/FORS1 & 13-09-1999  & 0.7--1.2  & 1.03--1.24 & photometric & visitor \\
2/FORS2 & 05-07-2000  &   0.5     &    1.11    & photometric & service \\
2/FORS2 & 27-09-2000  & 0.5--0.7  & 1.44--1.80 & photometric & service \\
2/FORS2 & 28-09-2000  & 0.9--1.0  & 1.22--2.27 & photometric & service \\
2/FORS2 & 29-09-2000  & 0.5--0.7  & 1.51--1.93 & photometric & service \\
\hline
\end{tabular}
\end{center}
Note: seeing values are given from the DIMM or the Guide Probe, as available 
in the observations log.
\label{tab_obslog}
\end{table*}

The MOS system allows positioning simultaneously $19$ movable slit
blade pairs on pre-selected objects of the CL~0048-2942 field, which
had to be previously imaged with the VLT. Seven different masks were
thus constructed and used from one to a maximum of four times
according to priority and brightness of the targeted objects. Each
single exposure lasted $30$ minutes. Due to the rigidity of the
MOS system, to the obvious clustering of targets in the field and to a
slightly different size of the f.o.v. between the cameras of the
$3.6$m telescope and the VLT, two or three slits of each mask had to
be positioned on objects that were not present in the photometric
catalogue (i.e.  fainter objects or objects just outside the smaller
f.o.v. of the EFOSC2 $3.6$m telescope camera).

All data were reduced according to standard procedures using
subroutines in the IRAF package.  Following bias subtraction and
flat-fielding, the spectra were extracted using the {\em apall} task
and optimal mode.  This task extracts a one-dimensional spectrum from
the 2D image and at the same time does the sky subtraction.
Wavelength calibration was carried out using Helium-Argon-HgCd lamps
in frames taken on the nights of the cluster observations and with
exactly the same setup and masks used on the objects. This process was
carried out using IRAF routines {\em identify}, {\em reidentify} and
{\em dispcor}, and the final rms error in the wavelength solution was
always less than 0.1~\AA\ with a maximum deviation of $0.15$ \AA\ for
all lines used.
Spectro-photometric standards were observed every night through a
long-slit with the same setup (grism, resolution, etc.) as the one used for
the science spectra. These allowed us to perform flux calibrations
that were carried out with standard IRAF tasks {\em standard} and
{\em sensfunc} with the standard stars and then {\em calibrate} with
the target galaxies. The resulting flux calibration accuracy is of the 
order of 10\% based on the rms given by {\em sensfunc}.

The spectra have been corrected for the interstellar reddening using
the Howarth (1983) Galactic reddening law; the 
value of $E(B-V) = 0.0141$ for the cluster
line-of-sight was calculated using the maps of dust IR emission from
Schlegel et al. (1998).
No correction regarding telluric absorption was made, because no atmospheric
stars were obtained at the time of the observations. However, this is 
not a problem as the contaminated regions of the spectra were 
removed from our analysis (see Section 5.2). 

Finally, all wavelength and flux calibrated sky subtracted spectra of
the same object were median combined. This strategy led to total
exposure times varying from $30$ minutes up to $5$ hours depending on
the object (some of them were targeted in different masks, which in
turn were used more than once). This allowed us to enhance the S/N of
faint objects and remove most cosmic rays.
In this combination procedure we sometimes discarded
single exposures of the fainter objects coming from masks observed
with poorer seeing and/or higher airmass.

The signal to noise ratio achieved on these final spectra is estimated
between 3 and 10 per Angstrom.

Redshifts were then measured from these final resulting spectra as
described in the following section.

\section{Redshift determination and galaxy catalogue}\label{sec_zcat}

For all galaxies with clear absorption lines, redshifts were measured
according to the Tonry \& Davis (1979) technique using the {\em
  rvsao.xcsao} package in IRAF that cross-correlates the galaxy
spectra with velocity template spectra. We used as velocity templates
$10$ spectra obtained, during the same runs (and with same setup as
that for spectrophotometric standards - see
section~\ref{sec_spectro}), for a sample of velocity standard stars,
as well as a spectrum of M31 (frequently used by F.~Durret) and $6$ other
templates kindly provided by C.~Adami, all taken with wavelength
resolutions comparable to that of the present data. Therefore a total
of $17$ template spectra were used for each galaxy redshift estimate.
The final redshift given in Table~\ref{tab_redshifts} is that
corresponding to the highest value of the Tonry \& Davis parameter
{\em R}, and the corresponding error bar is that given by {\em xcsao}.
In all cases, the final redshift was that obtained simultaneously with
at least several templates.

When emission lines were present, their positions were measured with a
Gaussian fit using the {\em splot.onedspec} package.  In order to
estimate error bars, when several emission lines were present, lines
were also measured individually in each spectrum of a given galaxy,
then the measured velocities were averaged and the dispersion of the
measurements was taken as the error bar. Nevertheless, in many cases, only
one emission line was detected, and we assumed it was
[OII]$\lambda$3727 since, in most cases, this hypothesis was supported
by the presence of a continuum strongly decreasing bluewards of this
line. In this case, the error bar on the emission line velocity should
formally be taken as that in the wavelength calibration quoted above.
However, when the [OII]$\lambda$3727 line is measurable on individual
spectra of a given galaxy, the dispersion is rather one order of
magnitude higher, giving typical error bars on the redshift of 0.0004,
comparable to absorption redshift error bars.


From a total of $66$ objects observed, we managed to extract reliable
spectra and determine a redshift measure for $54$ of them. A catalogue
of these objects is given in Table~\ref{tab_redshifts}, containing the
following columns:
\begin{itemize}
\item []$1$-- Galaxy identification number (as in the photometric
  catalogue, except for galaxies with running number larger than
  $900$, which were not present in the photometric catalogue - see
  section \ref{sec_spectro}).
\item []$2$ and $3$ -- Galaxy coordinates (equinox 2000.0) astrometrically
  corrected with software {\em Gaia} applied to the VLT image. The
  astrometric solution is accurate to $\sim 0.28''$ in right ascension
  and to $\sim 0.26''$ in declination.
\item []$4$ and $5$ -- Absorption line redshift and corresponding error.
\item []$6$-- Tonry \& Davis parameter ({\em R}).
\item []$7$ and $8$ -- Emission line redshift and corresponding error.
\item []$9$ -- Number of emission lines used to estimate the emission line
  redshift.
\item []$10$ and $11$ -- I-band aperture magnitude within a 4.5 arcsec
  aperture and respective error, as obtained with {\em SExtractor} on the
  $3.6$m images (see Andreon et al. 2004).
\item []$12$ -- This last column contains alert flags. {\em b}: blended
  objects in photometry though position is accurate in VLT image; {\em
    o}: objects outside the $3.6$m f.o.v. or at the image boundary;
  {\em t}: objects with lower signal-to-noise photometry (as can be
  deduced from the corresponding magnitude errors) due to the fact
  that only one of the several exposures obtained at the $3.6$m
  telescope covered the region occupied by the object in question; for
  these objects, the magnitude obtained is a rough estimate.
\end{itemize}

\begin{table*}[ht]
\caption{Catalogue for all galaxies with measured redshift in the field of 
CL~0048-2942 (see text for detail on the table entries).}
\begin{center}
\begin{tabular}{rrrrrrrrrrrr}
\hline
Galaxy &  $\alpha$    &	$\delta$	  & z$_{abs}$ & error & {\em R} & z$_{em}$ & error &N$_{em}$ & I & $\Delta$I & flag \\ 
id \#       & (J2000.0)  & (J2000.0)     & & z$_{abs}$ & & & z$_{em}$ &  &  &  & \\
\hline     
35  & 00 48 28.144 & -29 39 48.40 &        &        &      & 0.6382 &        & 1 & 21.5288 & 0.1091 & t \\
37  & 00 48 34.210 & -29 39 51.21 & 0.6387 & 0.0006 & 1.86 & 0.6390 &        & 1 & 21.0832 & 0.1091 & t \\
44  & 00 48 24.402 & -29 39 53.51 & 0.6962 & 0.0005 & 2.64 &        &	     &   & 21.5094 & 0.1091 & t \\
58  & 00 48 21.954 & -29 40 09.29 & 0.8215 & 0.0006 & 2.34 & 0.8224 &        & 1 & 20.9068 & 0.0205 &   \\
76  & 00 48 40.976 & -29 40 27.01 &        &        &      & 0.3970 & 0.0004 & 4 & 20.9160 & 0.0207 &   \\
102 & 00 48 36.953 & -29 40 51.86 & 0.6349 & 0.0004 & 3.78 &	    &	     &   & 21.2419 & 0.0278 &   \\
121 & 00 48 22.333 & -29 41 00.64 & 0.2246 & 0.0004 & 2.75 &	    &	     &   & 20.0203 & 0.0092 &   \\
128 & 00 48 29.390 & -29 40 51.45 & 0.6439 & 0.0004 & 3.84 &	    &	     &   & 20.7153 & 0.0172 &   \\
161 & 00 48 27.885 & -29 41 34.86 & 0.7966 & 0.0002 & 3.22 &	    & 	     &   & 21.2002 & 0.0267 &   \\
184 & 00 48 35.836 & -29 41 50.42 & 0.6364 & 0.0005 & 3.78 &	    &	     &   & 19.6551 & 0.0067 &   \\ 
188 & 00 48 26.940 & -29 41 49.71 &        &        &      & 0.2120 & 0.0002 & 5 & 20.7089 & 0.0171 &   \\
189 & 00 48 32.270 & -29 41 50.52 & 0.6422 & 0.0002 & 2.75 & 	    &	     &   & 21.4267 & 0.0329 &   \\
196 & 00 48 23.879 & -29 41 52.75 & 0.6994 & 0.0007 & 1.65 & 0.6989 &        & 1 & 21.4933 & 0.0349 &   \\
199 & 00 48 43.198 & -29 41 59.92 & 0.7210 & 0.0010 & 2.41 & 0.7165 &        & 1 & 19.7774 & 0.0074 & b \\
203 & 00 48 41.254 & -29 41 59.87 & 0.4605 & 0.0002 & 2.33 & 0.4619 &        & 1 & 20.8357 & 0.0192 &   \\
212 & 00 48 21.910 & -29 41 55.06 &        &        &      & 0.6010 & 0.0002 & 3 & 20.4501 & 0.0135 &   \\
213 & 00 48 32.725 & -29 42 04.07 & 0.6384 & 0.0001 & 5.41 &	    &	     &   & 21.0950 & 0.0243 & b \\
216 & 00 48 32.097 & -29 42 07.14 & 0.6384 & 0.0003 & 5.13 &	    &	     &   & 20.2979 & 0.0118 & b \\
217 & 00 48 31.423 & -29 42 10.19 & 0.6167 & 0.0004 & 2.89 &	    &	     &   & 19.7061 & 0.0070 & b \\
218 & 00 48 30.075 & -29 42 11.00 & 0.6502 & 0.0002 & 3.12 &	    &	     &   & 21.2275 & 0.0274 &   \\
231 & 00 48 26.213 & -29 42 17.65 & 0.6417 & 0.0007 & 1.33 &	    &	     &   & 21.1841 & 0.0264 & b \\
235 & 00 48 30.934 & -29 42 08.88 & 0.6362 & 0.0003 & 4.96 &	    &	     &   & 20.6515 & 0.0162 &   \\
244 & 00 48 35.483 & -29 42 23.98 & 0.4740 & 0.0002 & 7.26 &	    &	     &   & 21.2334 & 0.0276 &   \\
245 & 00 48 35.760 & -29 42 26.10 & 0.4720 & 0.0008 & 2.20 & 0.4715 &        & 1 & 19.9062 & 0.0083 &   \\
248 & 00 48 39.410 & -29 42 26.97 & 0.4611 & 0.0002 & 2.59 & 0.4603 & 0.0004 & 2 & 20.1946 & 0.0108 &   \\
265 & 00 48 20.162 & -29 42 31.01 & 0.2229 & 0.0002 & 9.54 &	    &	     &   & 20.4458 & 0.1091 & t \\
266 & 00 48 26.044 & -29 42 31.30 & 0.8284 & 0.0005 & 3.03 & 0.8290 &        & 1 & 20.8273 & 0.0190 &   \\
280 & 00 48 35.790 & -29 42 43.79 & 0.4733 & 0.0002 & 6.59 &        &        &   & 21.1794 & 0.0263 &   \\ 
286 & 00 48 20.724 & -29 42 44.52 & 0.6079 & 0.0005 & 2.72 &	    &	     &   & 21.3677 & 0.1091 & t \\
296 & 00 48 42.518 & -29 42 55.58 & 0.6320 & 0.0008 & 1.35 & 0.6300 &        & 1 & 20.7723 & 0.0181 &   \\
304 & 00 48 24.672 & -29 42 55.49 &        &        &      & 0.3529 &        & 1 & 21.3487 & 0.0306 &   \\
322 & 00 48 38.134 & -29 43 08.27 & 0.6366 & 0.0004 & 4.40 & 0.6364 &        & 1 & 20.8491 & 0.0194 &   \\
343 & 00 48 30.828 & -29 43 16.58 & 0.6353 & 0.0006 & 2.53 & 0.6345 &        & 1 & 20.2195 & 0.0110 &   \\
346 & 00 48 41.553 & -29 43 23.70 & 0.6997 & 0.0003 & 4.33 & 	    &	     &   & 20.7813 & 0.0183 &   \\
391 & 00 48 31.379 & -29 43 45.19 & 0.6287 & 0.0001 & 3.46 & 0.6289 &        & 1 & 21.0323 & 0.0230 &   \\
396 & 00 48 39.315 & -29 43 52.93 & 0.6415 & 0.0004 & 3.11 & 	    &	     &   & 20.6534 & 0.0163 &   \\
399 & 00 48 38.531 & -29 43 54.94 & 0.6757 & 0.0004 & 4.20 & 0.6755 &        & 1 & 20.6518 & 0.0162 &   \\
400 & 00 48 41.942 & -29 43 55.86 & 0.6091 & 0.0007 & 1.53 & 0.6080 &        & 1 & 21.8393 & 0.0479 &   \\
404 & 00 48 37.741 & -29 44 00.29 & 0.6345 & 0.0006 & 1.96 & 0.6345 &        & 1 & 20.9757 & 0.0218 &   \\
419 & 00 48 33.234 & -29 44 24.83 & 0.5461 & 0.0002 & 2.93 & 0.5475 &        & 1 & 20.9346 & 0.0210 & b \\
421 & 00 48 35.362 & -29 44 17.75 & 0.6367 & 0.0003 & 2.81 & 0.6352 &        & 1 & 21.0307 & 0.0229 &   \\
424 & 00 48 42.717 & -29 44 21.03 & 0.7978 & 0.0003 &      &        &        &   & 21.9294 & 0.0521 &   \\
428 & 00 48 22.022 & -29 44 18.39 & 0.5403 & 0.0004 & 3.70 & 	    &	     &   & 20.6207 & 0.0158 &   \\
439 & 00 48 42.702 & -29 44 29.34 & 	   &        &      & 0.7968 &        & 1 & 22.0275 & 0.0570 &   \\
443 & 00 48 28.492 & -29 39 29.77 &        &        &      & 0.6405 &        & 1 & 21.3254 & 0.1091 & t \\
449 & 00 48 21.537 & -29 39 33.32 & 0.6351 & 0.0002 & 1.26 & 0.6370 &        & 1 & 21.6434 & 0.2688 & t+b \\
451 & 00 48 22.202 & -29 39 36.44 & 0.6356 & 0.0003 & 4.96 & 	    &	     &   & 20.2855 & 0.0478 & t+b \\
452 & 00 48 21.908 & -29 39 35.97 & 0.6404 & 0.0002 & 1.22 & 0.6369 &        & 1 & 21.4085 & 0.1091 & t+b \\
901 & 00 48 43.774 & -29 44 52.00 &        &        &      & 0.5193 & 	     & 1 &         &        & o \\
902 & 00 48 46.147 & -29 42 14.80 & 0.6302 & 0.0005 & 2.80 &        &        &   &         &        & o \\
903 & 00 48 44.169 & -29 40 45.16 & 0.5291 & 0.0004 & 2.88 &        &	     &   &         &        & o \\
906 & 00 48 42.497 & -29 44 57.53 & 0.5912 & 0.0005 & 3.09 &	    &	     &   &         &        & o \\
907 & 00 48 31.404 & -29 42 07.67 &        &        &      & 0.2988 & 0.0007 & 5 &         &        & b \\
908 & 00 48 41.564 & -29 39 06.40 & 0.3368 & 0.0005 & 2.59 &	    &	     &   &         &        & o \\
\hline
\end{tabular}
\label{tab_redshifts}
\end{center}
\end{table*}

None of the $54$ objects catalogued in Table~\ref{tab_redshifts}
turned out to be stars, a result of the field being close to the
direction of the south galactic pole allied to a cautious target
selection.

\begin{figure}
\centering
\includegraphics[width=\textwidth,]{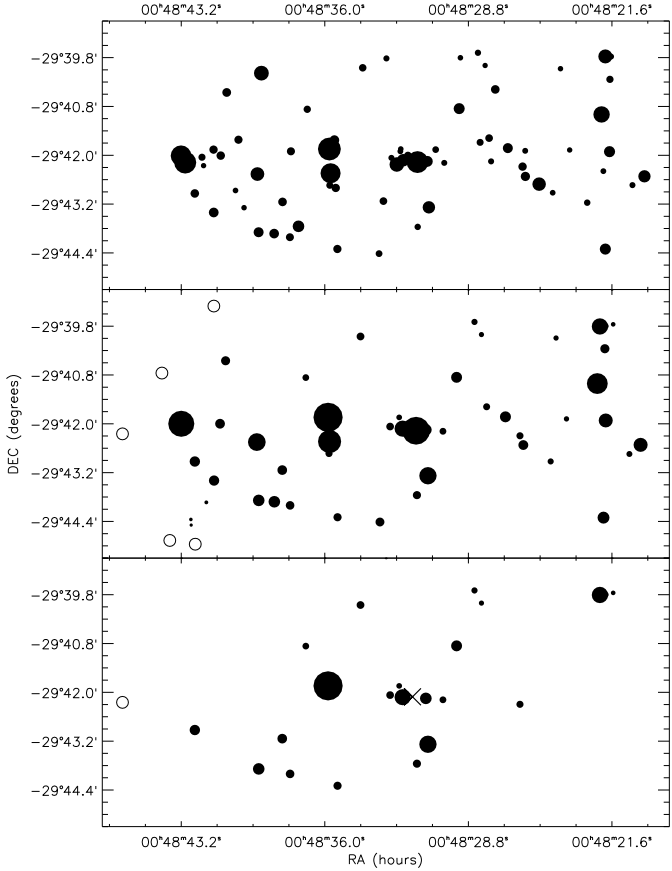}
\caption{The uppermost panel displays the positions of the $73$ objects in 
  the cluster line-of-sight that were originally selected for
  spectroscopy (see text).  Symbol size is proportional to
  each galaxy's flux in the I-band (as computed from the aperture
  magnitude referred to in section \ref{sec_phot}).  The middle panel
  shows the positions of the $54$ galaxies for which spectra was
  obtained and a reliable redshift was measured - listed in Table
  \ref{tab_redshifts}. Symbol size is, again, proportional to each
  galaxy's flux in the I-band (and is listed in column 10 of
  Table~\ref{tab_redshifts}). Galaxies with no magnitude measure
  available (i.e. with identification number larger than $900$) are
  noted by an open symbol (size, in these cases, is arbitrary). The
  lower panel maps the positions of the $23$ galaxies belonging to the
  cluster (see section~\ref{sec_clus}). The cluster centre, as defined
  by the cluster detection algorithm, is identified by the cross.
}
\label{ALLobjs}
\end{figure}

Completeness of our spectroscopic observations is hard to estimate
since we ended up targeting some (few) objects outside the original
f.o.v. and others outside our initial magnitude range of $20.0 <$ I
$<22.0$.  The upper and middle panels of Figures \ref{ALLobjs} show,
respectively, the $73$ objects of the photometric catalogue in the
cluster field originally selected for spectroscopy (ie. within the
originally defined magnitude range), and the $54$ galaxies with
redshift measures (provided in Table \ref{tab_redshifts}).

From the $73$ objects originally elected for spectroscopic follow-up,
$44$ are present in our redshift catalogue. These figures give a rough
indication of a spectroscopic completeness of about 60\% for the
adopted magnitude range.

\section{Cluster members - definition of the sample and some structural and 
dynamical properties}\label{sec_clus}

\begin{figure}
\centering
\includegraphics[width=7cm]{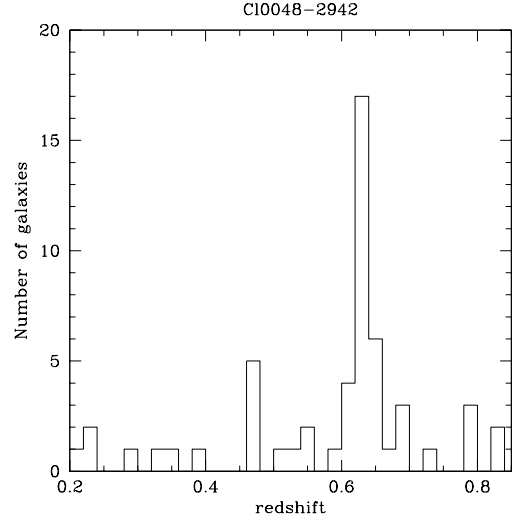}
\caption{Histogram of redshifts for all $54$ galaxies 
in the cluster field with a line-of-sight velocity measure.}
\label{hist_z}
\end{figure}

The redshift histogram for all galaxies of Table~\ref{tab_redshifts}
is shown in Fig.~\ref{hist_z}, where the conspicuous spike denotes the
presence of the cluster.

In order to define the redshift interval corresponding to the cluster,
we first converted redshift values into velocities. In this step, we
note that whenever an object had two measures, z$_{abs}$ and z$_{em}$,
we took the best one considering the quality of the absorption
measure (as given by the Tonry \& Davis {\em R} parameter).
We then applied the ``velocity gap'' method described e.g. by Katgert
et al. (1996) for identifying gaps in the redshift distribution larger
than a certain value; this gap should separate outliers from the
limits of the cluster along the line-of-sight. This method does not
impose a Gaussian distribution on the final velocity distribution,
which can be an advantage when the system gives no evident sign of
being relaxed/virialised (see below). We adopted a velocity gap of
$1500$ km/s as in Benoist et al. (2002), since we have very similar
data. We thus defined the cluster as the bulk of galaxies with
redshifts comprised in the interval $[0.6287-0.6502]$. Based on this
result, we shall hereafter consider the value $0.64$ as the mean
redshift of cluster CL~0048-2942.

On the upper limit of this redshift interval, galaxy $\#218$ (taken as
the cluster member with the highest redshift value) remains somewhat
isolated (with a difference of $\sim 900$ km/s to the immediately
closest cluster neighbour in velocity space along the line-of-sight).  
This could simply
be due to the fact that our redshift catalogue is incomplete. On the
other hand, tightening up the criterion used in the gap-method,
"ejects" this galaxy from the cluster once we replace the fixed gap of
$1500$ km/s by the velocity dispersion (either standard or bi-weight)
of the group individualised at each step (an alternative choice for
the gap used by other authors). In doubt, we keep galaxy $\#218$ in
the cluster, as a marginally bound member. We have tested that this
won't affect any of the results presented in this paper.

On the other ``side'', right below the lower limit of our redshift
interval, lies galaxy $\#217$. This is a previously known
radio-galaxy, source NVSS J004831-294207 (Condon et al. 1998), for
which Brown, Webster \& Boyle (2001) had computed a photometric
redshift of $0.542$, and that we had originally considered as a strong
candidate to be the brightest cluster member (BCM), given (1) its
proximity to the centre determined by the cluster detection algorithm,
and the fact that it is the brightest galaxy nearby; (2) the
well-known fact that these galaxies act as signposts for density
enhancements (e.g.  Eales 1985; Yates, Miller \& Peacock 1989; Deltorn
et al. 1997; Best et al.  2003 and references therein). However, the
gap of $\simeq -1770$ km/s between this galaxy and the lowest-z
cluster member, according to our previous definition, is much too
large for us to include it in the cluster: galaxy $\#217$ is probably
merely close to the cluster in projection.

A total of $23$ cluster member galaxies are thus identified, with a
spatial distribution shown in the bottom panel of Fig.~\ref{ALLobjs}.
We note that, at the cluster mean redshift of $\sim 0.64$, the square
f.o.v.  of size approximately $5.4$ arcmin long enclosing the $23$
cluster galaxies subtends $\sim 2.2$ Mpc ($\Omega_\Lambda=0.7$,
$\Omega_m=0.3$ and $H_0=70$ km s$^{-1}$ Mpc$^{-1}$), and corresponds
to the typical virial ``diameter'' of a cluster (e.g. Lima Neto et al.
2003 and references therein). Therefore, we expect to be probing the
cluster population till regions well outside the cluster core (even
though we miss the cluster infall region).

The distribution of galaxies displayed in the lower panel of
Fig.~\ref{ALLobjs} is somewhat filamentary. This overall shape - which
is definitely not a result of incomplete spectroscopy as can be seen
by consulting the two uppermost panels of the same figure - could
indicate a dynamically young age for our cluster (Lubin et al. 2002
and references therein).  The fact that the velocity histogram for the
$23$ cluster members is slightly skewed and non-Gaussian may be,
incompleteness apart, another indication that the system may not be
virialised. No particular substructure is visible in the spatial
galaxy distribution on the plane of the sky.  In an independent work
on CL~0048-2942, however, but probing roughly the same cluster region,
La Barbera et al. (2003) find that the spatial distribution of the
cluster galaxies shows a clumpy structure, with a main over-density of
radius $\sim 0.5$ Mpc, and at least two other clumps $\sim 1$~Mpc away
from the centre.

We have therefore defined the cluster as corresponding to the
$[0.6287-0.6502]$ redshift interval (or, equivalently, the
$[135~642-138~751]$ km/s line-of-sight velocity range). The cluster
line-of-sight velocity dispersion then becomes approximately $680$
$\pm$ 140 km/s ($625$ km/s if we use the bi-weight estimator), a value which is
within the lower range of dispersions derived for clusters observed at
comparable redshifts (e.g.  Valtchanov et al.  2003, Wittman et al.
2003, Tran et al. 2003) with equivalent sampling (i.e. number of
observed cluster members and selection for spectroscopy). This denotes
a mild richness for CL~0048-2942 according to the relations derived by
Yee \& Ellingson (2003).  In fact, Andreon et al. (2004), based solely
on photometric data, estimated an Abell (1958) richness class R$=0-1$
for this cluster, again consistent with the works of Yee and
collaborators (see also Yee \& L\'opez-Cruz 1999), although we do
refrain from ascertaining the validity of extending such
classifications to such high-z systems.

The corresponding X-ray temperature following the $\sigma _v - T_{\rm
  X}$ relation defined by Wu et al. (1998) would be $\sim$2.5 keV
(assuming the bi-weight velocity dispersion; $\sim$2.9 keV for the
standard one) and the corresponding X-ray luminosity would then be
$L_X \sim 1.3\ 10^{44}$ erg/s ($\sim 1.9\ 10^{44}$ erg/s) following
the Wu et al. (1999) relation, or $L_X \sim 9.2\ 10^{43}$ erg/s ($\sim
1.4\ 10^{44}$ erg/s) according to the Arnaud \& Evrard (1999)
relation; these values suggest that this is not a very hot cluster.
However, a ROSAT PSPC image, serendipitously containing CL~0048-2942,
shows an excess of counts (a factor 2.3) in the cluster area with
respect to the background, suggesting that we are indeed dealing with
an X-ray emitting cluster.  It could therefore be an interesting
target for the present generation X-ray satellites such as XMM-Newton
and/or Chandra. In fact, Lubin et al. (2004) do derive with XMM-Newton
a similarly modest X-ray emission for their z $=0.76$ and z $=0.90$
clusters, stressing that this seems to be a usual trend observed for
optically selected clusters.

\section{Stellar population synthesis}\label{sec_starpop}

Several works (e.g.  Poggianti et al. 1999, Kelson et al. 2000, 
Balogh et al. 2002, Tran
et al. 2003, Quintero et al. 2004, and references within these papers)
have shown that a significant fraction of galaxies in clusters seem to
have suffered a dramatic decrease of their star-formation rate over
the last $\sim 1-1.5$ Gyr which could have been preceded by an intense
burst of star formation (see Balogh et al. 1999 for references for and
against this debatable issue).  This stage is followed by a period of
passive evolution in luminosity. The percentage of the same type of
objects found in field surveys seems to be lower (see Tran et al. 2004
and references therein). In this paper we will use a population
synthesis analysis to determine the stellar content (or evolutionary
stage) of the cluster galaxies of CL~0048-2942 at different radial
distances from the cluster centre. Galaxies of the field sample,
analysed in the very same way, provide the term of comparison in order
to deduce the impact of environmental effects and dynamical
interactions typical of clusters on the evolution of their member
galaxies in what concerns, in particular, their stellar populations.

\subsection{The synthesis method}\label{metodo}

To compute the stellar content of our observed galaxies, we have performed a
stellar population synthesis by means of a mathematical algorithm
developed by Pelat (1997), which gives a unique solution (Global
Principal Geometrical solution or GPG solution), contrary to other
methods widely used for population synthesis.  It makes use of the
equivalent widths (EWs) of all the absorption features found in the
spectrum.  Essentially, it considers a galaxy as being made up of a
set of stars with different spectral types, luminosity classes and
metallicities. This particular composition will carry its own
signature in terms of the EWs of the absorption lines. The method thus
defines the galaxy composition by reproducing its signature as closely
as possible, taking into account all observed EWs in the spectrum of
the galaxy. This is done by matching the observed EWs of the galaxy
spectra with stellar EWs of a combination of stars; the best match is
chosen by minimising (by the least squares method) the following
equation:

\begin{displaymath}
{D}^{2}=\sum^{n_{\lambda}}_{j=1}(W_{Obs, j}-W_{syn,j})^{2}P_{j},\hspace{3mm} P_{j}\geq0
\end{displaymath}

\noindent
where {\em D} is the sum, for all absorption features, of the difference
between the observed EW ($W_{Obs, j}$, for line j) and the synthetic EW
($W_{syn,j}$, for line j). $P_{j}$ is the weight of line j (defining which 
lines are more important in terms of specific characteristics of one or
another stellar type) and

\begin{displaymath}
W_{syn,j}={\frac{\sum^{n_\star}_{i=1}(W_{ji}I_{ji}k_{i})}{\sum^{n_{\star}}_{i=1}(I_{ji}k_{i})}},
\hspace{5mm}{j=i,...,n_\lambda}
\end{displaymath}
with $W_{ji}$ the EW of line j in star i, $I_{ji}$ the value of the
continuum for line j in star i and $k_{i}$ the contribution of star i
to the total light at 4200~\AA\ (taken as the reference wavelength).

In order to ensure the physical validity of the solutions, the
normalisation and positivity constraints must be satisfied:

\begin{displaymath}\sum^{n\star}_{i=1}k_{i}=1\end{displaymath}

\begin{displaymath}k_{i}\geq 0, i=1,...,n\star \end{displaymath}

The accuracy of the fit is estimated through means of the {\em
  distance}, {\em D}. This value, as can be seen above, depends on the
number of absorption features used, and the smaller D is, the better
the fit.  In addition, residuals estimated over the continuum help us
to verify the accuracy of the solution found.  Also, the internal
reddening $E_{B-V}$ is a free parameter given by the method in an
indirect way, i.e., we match the continuum as closely as possible by
applying an internal extinction law (Howarth 1983; Cardelli 1989)
whenever needed.

The complete description of this method and all its theoretical
aspects can be found in Pelat (1997).

\subsection{The stellar database}

The stellar database used in this work was compiled from the stellar
library of Pickles (1998) which gathers $131$ stellar spectra of every
spectral type and luminosity class from $1150$ to $10620$ \AA\ with a
resolution of R $= 500$. The library also includes some metal-rich and
metal-poor stars.

A total of $37$ stars were chosen from this stellar library in order
to cover the temperature/gravity parameter space as much as possible
without being degenerate. In fact, in order to prevent stellar library
degeneracy, i.e.  having two different stellar types with spectral
energy distributions similar enough to be indistinguishable in a
mathematical sense, we cannot include as many stellar types as we
would like to.  Some low metallicity stars have also been included.
The wavelength used was $2500-5500$ \AA\ in order to match the
rest-frame data of our galaxies. Note that for cluster galaxies, our
spectroscopic setup provides coverage of the rest-frame band that goes
from approximately $2700$ to $5300$ \AA. Regarding field galaxies,
this is a compromise value since some of them, with lower and higher
values of redshift, will not be covered in the blue or in the red
part, respectively, by the stellar base. However, this is not a
problem since the number of features lost due to this issue is not
significant in comparison with all the features used in the synthesis
(see below).

The ages of all stars in our stellar library were estimated by
comparing their position in the HR diagram with the theoretical models
of Bressan et al. (1993). These authors have computed evolutionary
stellar tracks for several masses and different metallicities.  We
have used models from low mass stars of 0.6 M$\odot$ to massive stars
of 9 M$\odot$ and metallicities of Z $= 0.05$, Z $= 0.02$ and Z $=
0.008$.  We have plotted the stars along these tracks.  According to a
star's position in the HR diagram relative to the models we then
estimated its mass and age. This procedure is especially important for
main sequence stars because, as their life span is usually quite long
(except for very massive hot stars), stars in the beginning and in the
end of the main sequence will have very different ages.

Table~\ref{tab_estrelas} lists all stars of our stellar library
together with the ages calculated as explained above.

\begin{table}
\centering
\caption{Stars in the stellar database. The ages were calculated using
the Padova stellar tracks (see text).}
\label{tab_estrelas}
\begin{tabular}{|l|r|}
\noalign{\smallskip}
\hline
\noalign{\smallskip}
star&age (yr) \\
\noalign{\smallskip}
\hline
\noalign{\smallskip}
 B0V     &  $<$ 10$^7$ \\
 B5-7V   & 4.4 10$^7$ \\
 A0V     & 3.5 10$^8$ \\
 A5V     & 4.0 10$^8$ \\
 F2V     & 5.7 10$^8$ \\
 F8V     & 3.6 10$^{9}$ \\
 w F8V   &6.5 10$^{9}$ \\
 G5V     &1.3 10$^{10}$ \\
 w G5V   &4.3 10$^{9}$ \\
 K2V     &4.0 10$^{10}$ \\
 K4V     &5.8 10$^{10}$ \\
 M2V     &$>$ 10$^{10}$ \\
        & \\
 B1-2III & 2.8 10$^7$ \\
 B5III   & 6.5 - 10 10$^7$ \\
 A0III   & 3 10$^8$ \\
 A5III   & 5 - 6  10$^8$ \\
 F0III   & 8.8 10$^8$ \\
 F5III   & 1.2 - 2 10$^9$ \\
 G0III   & $\sim$ 10$^{9}$ \\
 G5III   & 3 - 4.5 10$^{9}$ \\
 w G5III & $\sim$ 10$^{9}$ \\
 w G8III & $\sim$ 10$^{9}$ \\
 K0III   & 5.3 10$^{9}$ \\
 K3III   & 6 - 8.4 10$^9$ \\
 r K3III & 7.5 10$^9$ \\
 w K3III &  $\sim$ 10$^9$ \\
 K5III   & 3.4 - 6.0 10$^9$ \\
 M5III   & 3.5 - 4.5 10$^9$ \\
         & \\
 B0I     &  7 10$^6$ \\
 B5I    &   9 10$^6$ \\
 A0I    &   1.3 10$^7$ \\
 A2I    &   1.3 10$^7$ \\
 F0I     &  1.8 10$^7$ \\
 F8I    &   2.5 10$^7$ \\
 G8I     &  1.5 10$^7$ \\
 K3I    &   1.5 10$^7$ \\
 M2I    &   4.8 10$^7$ \\
\noalign{\smallskip}
\hline
\end{tabular}
\end{table}

All absorption features (totaling $63$) present in the stellar
spectra were first identified and their wavelength interval defined,
taking into account the shape of the absorption features in both hot
and cool stars of the stellar library. Table~\ref{EW} gives the
features and the respective wavelength intervals defined.


\begin{table*}[ht]
\caption{\label{EW}
The 63 wavelength intervals defined for measuring the  equivalent 
widths (all values are in units of~\AA).}
\begin{center}
\begin{tabular}{lcc|lcc}
\hline
Line Identification & $\lambda_{\rm central}$ & Wavelength Interval&
Line Identification & $\lambda_{\rm central}$ & Wavelength Interval \\
\hline
MgI& 2517.0& 2500 --        2534         &          H9,CN L band,FeI,MgI,HeI& 3835.0& 3808 -- 3862\\      
FeII& 2550.0& 2534 --        2566         &	   	 H8,CN L band,FeI,SiI,HeI& 3885.0& 3862 -- 3908\\
FeII& 2590.5& 2566 --	 2615         &	   	 CaIIK& 3930.0& 3908 --	 3952                  \\
FeII,CrII,FeI& 2635.0& 2615 --	 2655         &	      	 CaIIH,H$\epsilon$& 3970.0& 3952 --	 3988  \\
FeII& 2672.5& 2655 --	 2690         &	   	 FeI,HeI& 4004.0& 3988 --	 4020          \\
FeII& 2710.0& 2690 --	 2730         &	   	 FeI,HeI& 4037.0& 4020 --	 4054          \\
FeII& 2751.5& 2730 --	 2773         &	   	 FeI,SrII& 4068.0& 4054 --	 4082          \\
FeII,MgII& 2798.0& 2773 --        2823         &	 H$\delta$& 4100.0& 4082 --	 4118          \\
MgI,FeII& 2842.5& 2823 --	 2862         &	   	 FeI& 4138.5& 4118 --	 4159                  \\
FeI,FeII& 2884.0& 2862 --	 2906         &	   	 CN& 4186.5& 4159 --	 4214                  \\
FeI,FeII,CrII& 2939.0& 2906 --	 2972         &	   	 CaI& 4229.0& 4214 --	 4244                  \\
FeI,FeII,CrI,CrII& 2982.0& 2972 --	 2992 &	   	 FeI,CrI& 4260.5& 4244 --4277          \\
FeI,MnI,CrI,CrII& 3009.0& 2992 --	 3026 &	   	 CH G band,FeI,CrI& 4297.5& 4277 -- 4318  \\
FeI,OH,TiII& 3052.5& 3026 --	 3079         &	   	 H$\gamma$,FeI,FeII& 4341.0& 4318 --	 4364  \\
CoI,FeI,OH,& 3096.0& 3079 --	 3113         & 	 FeI,C$_2$,FeII,TiII& 4392.0& 4364 --	 4420  \\
OH,FeI,CrII,CH& 3122.5& 3113 --	 3132         &	   	 FeI,CaI,TiO,MgII,HeI& 4446.0& 4420 --	 4472  \\
FeI,CH& 3139.5& 3132 --	 3147         &	   	 CH,CN& 4489.0& 4472 --	 4506                  \\
OH,FeI,FeII,CH,TiII& 3178.0& 3147 --	 3209 &	         FeI,FeII,TiII& 4537.0& 4506 --	 4568     \\
FeI,TiII,MnI& 3236.5& 3209 --	 3264         &	         FeI,FeII,TiO,TiII,CN,CaI& 4595.0& 4568 -- 4622\\
FeI& 3281.0& 3264 --	 3298         &	    	 FeI,TiO,C$_2$& 4655.0& 4622 --	 4688          \\
FeI,TiII& 3323.5& 3298 --	 3349         &	         FeI,MgI,TiI,HeI,NiI,C$_2$& 4714.5&4688 -- 4741\\
NH,FeI& 3369.0& 3349 --	 3389         &	    	 FeI,MgH,NiI,MnI,TiO& 4771.5& 4741 --	 4802  \\
FeI,NiI& 3407.5& 3389 --	 3426         &	   	 TiO,MgH,CN,MnI& 4818.5& 4802 --	 4835  \\
FeI,NiI,CoI& 3443.0& 3426 --	 3460         &	       	 H$\beta$,TiO,FeI& 4857.0& 4835 --	 4879  \\
FeI,NiI,MnII& 3482.0& 3460 --	 3504         &	   	 FeI& 4889.0& 4879 --	 4899                  \\
FeI,NiI& 3523.5& 3504 --	 3543         &	   	 FeI,FeII,CN,HeI& 4922.5& 4899 --	 4946  \\
FeI,NiI,CrI& 3570.5& 3543 -- 3598     &	   	 FeI,TiO,TiI& 4972.0& 4946 --	 4998          \\
FeI,NiI,CaI,TiI& 3629.5& 3598 -- 3661 &  	 FeI,FeII,TiO,CN,HeI,TiI& 5028.0& 4998 -- 5058 \\
FeI,CrI,TiI,NiI& 3677.0& 3661 -- 3693 &  	 FeI& 5107.0& 5058 --	 5156                  \\
FeI& 3717.0& 3693 --	 3741         &	   	 FeI,MgI+MgH& 5198.0& 5156 --	 5240          \\
FeI,TiII,CN& 3762.5& 3741 --	 3784         &	   	 FeI& 5274.0& 5240 --	 5308                  \\
H10,CN L band& 3796.0& 3784 --	 3808 &  &  & \\
\hline
\end{tabular}
\end{center}
\vspace{-5mm}
\end{table*}


The equivalent widths of all these features were then measured, as
well as the values of the continuum for each feature. The continuum
level has been determined globally over the whole wavelength
range. The error due to the uncertainty on the continuum level is
dominant over all other measurements and statistical uncertainties.
For strong well defined stellar features (e.g.  CaII H, K) this error
is always less than or equal to $1$~\AA\ in absolute value. It can
however reach a few Angstr\"oms for wide bands and strong blends in
the stars.

The EWs of the same spectral features have been measured in the
spectra of the observed galaxies.  For these, we discarded the
intervals corresponding to any emission lines present, as well as
those showing atmospheric absorption features.  We do stress the fact
that discarding some intervals of the wavelength coverage does not
affect the results of the method, as we still use a very large number
of parameters (see $n_f$ in Tabs.~\ref{tab_0048} and \ref{tab_campo}).

\subsection{Results}\label{sec_res}

We have synthesised the $54$ spectra available. However, for $10$ of
them no reliable solution could be found (e.g. too large values of
{\em D}, shape of the continuum not matching the observed spectra,
etc).  Of the $44$ objects for which a solution was achieved (in
terms of equivalent widths as well as of the continuum), $19$ are
cluster members, belonging to the cluster redshift range
$[0.6287-0.6502]$, while the remaining $25$ are field
galaxies ($17$ have z $<0.6287$ and $8$ have z $>0.6502$).  

We further note that the population synthesis method nicely provided a
totally independent confirmation of the redshift value obtained for
each object. This is accomplished because the synthetic galactic
spectra, made of a combination of stellar spectra (i.e. with z = 0),
must always match perfectly the observed galactic spectra in terms of
spectral features.

Due to the nature of the synthesis method we use, it is important to
have in mind several considerations when analysing the results
obtained for each object. In particular, we do not pretend to derive
the stellar content of each galaxy in its very details, but to have an
idea of its main components in terms of spectral type and age; note that 
the young population is given not only by hot stars, but also by supergiants.

Given these considerations, we have divided the stars 
into three groups according
to their age: young stars (with ages between 10$^6$ and 10$^8$ years),
intermediate ones (with ages around 10$^9$ years) and old stars (with
ages greater than 10$^{10}$ years).  These different age sequences are
found within each luminosity class, except for the supergiants where
all stars are young. Also, no old stars are found within the giants.
So, six different groups were defined in order to characterize the
stellar populations obtained for the galactic spectra: main sequence
young, intermediate and old populations; giant young and intermediate
populations and finally supergiants (all young stars).

\begin{figure*}
\centering
\includegraphics[width=\textwidth]{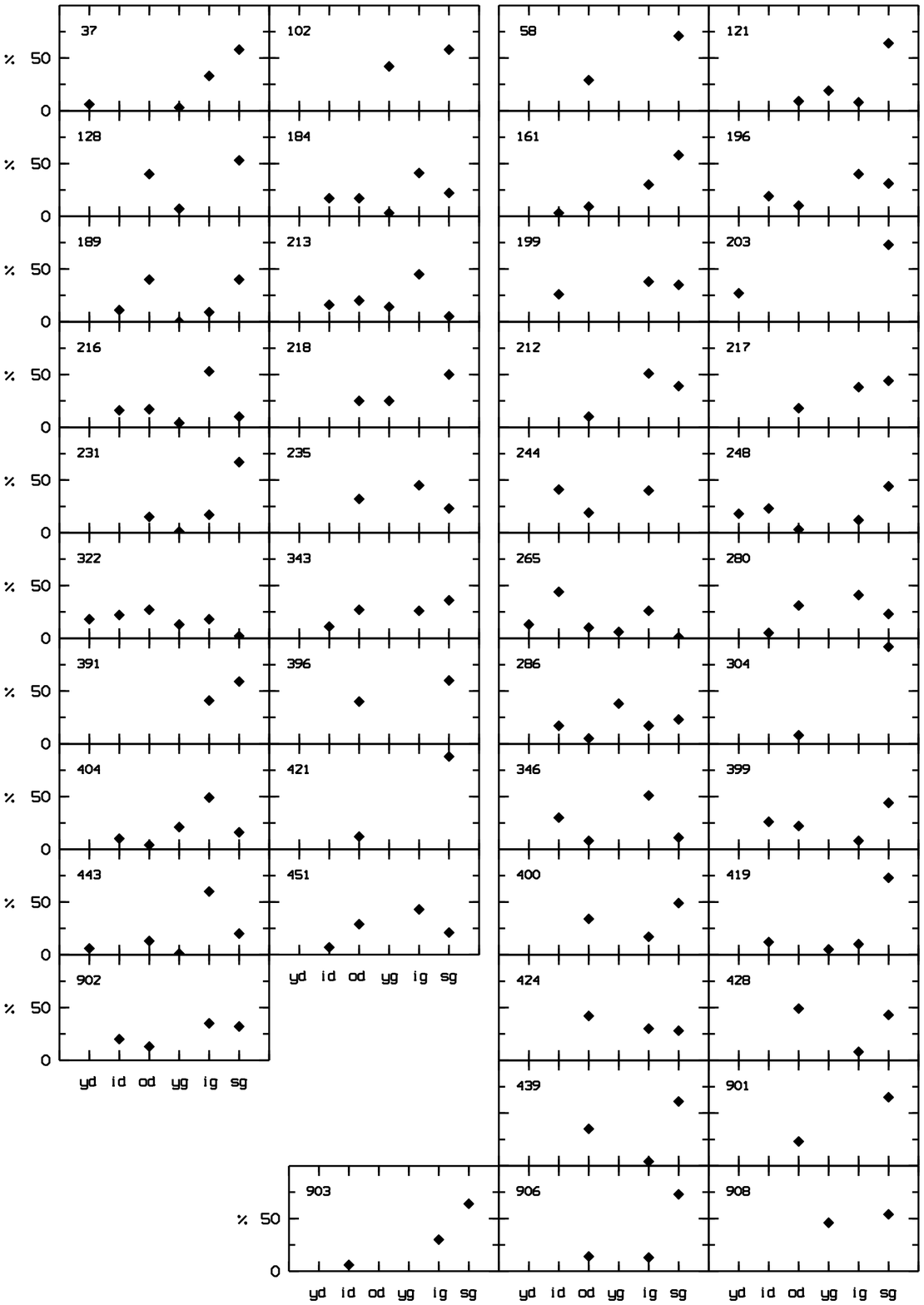}
\caption{Stellar population found for the 44 objects, 19 in the cluster (left
panels) and 25 in the field (right panels). In abcissae we have
plotted the 6 population components defined (see text): yd - young
dwarfs, id - intermediate dwarfs, od - old dwarfs, yg - young giants,
ig - intermediate giants and sg - supergiants.}  \label{sintese}
\end{figure*}

Tables~\ref{tab_0048} and \ref{tab_campo} and Figure~\ref{sintese} 
show the populations
obtained with our synthesis code for all 44 galaxies - 19 in the
cluster and 25 in the field - taking into account the six population
components we defined.  We also indicate in the Tables, for each galaxy, 
the value
of D and the number of spectral features used, $n_f$, which together
give an estimate of the reliability of the solution (see
section~\ref{sec_starpop}). This solution is, in each case, the best
match for the spectrum in question.

However, as one can see by inspection of Figure~\ref{sintese} and
 Tables~\ref{tab_0048} and
\ref{tab_campo}, the quality of the solutions is not the same for
all objects. In fact, and even taking into account the value of
$n_f$, some parameters D are quite big, denoting a not so good
solution. This is due to several factors, mainly the difficulty of
positioning the continuum, the existence of more or less emission
lines, the places where the atmospheric features fall within the
spectra (in relation, of course, to the redshift of each object) and
above all the existence or not of a good number of really well defined
absorption features.

Of all solutions we present, the ones for galaxies \#102, \#451 and
\#901 must be viewed with extra caution, as their values of D are
really quite large, denoting a solution of poor quality in terms of
the EWs. We have, however, chosen to present them since no better
solutions could be found for these galaxies and because the continuum
fit is quite reasonable.  For the remaining galaxies, even if some
still have values of D of the order of 150, most of them have this
parameter around 100 or smaller (several around 20) which denotes very
good fits in terms of the EWs and also in terms of the continuum
(linked, in fact, to the EWs).

Tables~\ref{tab_0048} and \ref{tab_campo} also list E(B-V), 
the internal reddening of each object, which
is a measure of the dust content
within the galaxy.  We can see by inspection of the these tables
that only small quantities of dust seem to be
present in both cluster and field galaxies. However, a degeneracy
exists between the internal dust of the galaxy and the blue continuum
of the spectrum, i.e. we can fit the observed continuum either with a
given amount of dust plus a given blue slope, or just with a certain
amount of dust (smaller in this case). This happens because both
contributions have the effect of blueing the spectrum. In this work we
have synthesised all spectra in such a way that the continuum is well
fit by the stars and by the dust (defined by a certain amount of
$E(B-V)$), thus assuming that the continuum entirely originates 
from the stellar content of the host galaxy and from the dust (if
present).  Instead, we could have found a blue excess in the continuum
by adding more dust to the solutions and then argued about this
quantity coming from nebular emission due to photoionization by hot
stars (since some B-type stars appear in the solution of the synthesis
this is quite normal).  However, in this case it would be difficult to
quantify both the quantity of dust present in the object and the slope
of the blue continuum.  That is why we chose not to include this
contribution.  However, we point to the fact that larger amounts of
dust can indeed be present.  In order to disentangle these two effects
we would need observations in the infrared.


\begin{table*}[ht]
{
\caption{Stellar populations for the cluster galaxies in terms of the
six stellar components defined (see text): main sequence (luminosity class V)
young, intermediate and old populations; giant (luminosity class III) young
and intermediate populations; young supergiant (luminosity class I) 
populations.
All values are in percentage and give the contribution of each 
population component to the total light at $\lambda_{ref} = 4200 \AA$.
The distance D, the number of features ($n_f$) (see section~\ref{metodo}) used 
for each solution found and the value of E(B-V), when applicable, 
are also given. The relative position of each galaxy within the cluster (see Fig.~\ref{psg} is indicated in the second row (``ct'' stands for 
centre or core; ``it'' denotes the intermediate region and ``p'' identifies 
peripheral galaxies).  
}
\label{tab_0048}
\centerline{
\begin{tabular}{rccccccc|rcc}
\noalign{\smallskip}
\hline
\noalign{\smallskip}
CL0048&position&V young&V int.&V old&III young&III int.& I young&D&n$_f$&E(B-V)\\
\noalign{\smallskip}
\hline
\noalign{\smallskip}
218&ct.&  &  &25&  &25&50&193&36&0.0\\
235&ct.&  &  &32&  &45&23&63 &33&0.0\\
189&ct.&  &11&40&  & 9&40&106&30&0.0\\
213&ct.&  &16&20&14&45& 5&75 &37&0.0\\
216&ct.&  &16&17& 4&53&10&125&43&0.0\\
231&it.&  &  &15& 1&17&67&50 &18&0.0\\
102&it.&  &  &  &  &42&58&582&33&0.0\\
128&it.&  &  &40&  & 7&53&96 &32&0.0\\
184&it.&  &17&17& 3&41&22&56 &35&0.0\\
322&it.&18&22&27&13&18& 2&36 &33&0.0\\
343&it.&  &11&27&  &26&36&117&41&0.0\\
391&it.&  &  &  &  &41&59&75 &19&0.0\\
396&p. &  &  &40&  &  &60&52 &19&0.0\\
404&p. &  &10&4 &21&49&16&60 &32&0.15\\
421&p. &  &  &12&  &  &88&164&26&0.0\\
443&p. & 6&  &13& 1&60&20&66 &28&0.0\\
451&p. &  &7 &29&  &43&21&280&31&0.0\\
902&p. &  &20&13&  &35&32&156&35&0.0\\
37 &p. & 6&  &  & 3&33&58&17 &19&0.0\\
\noalign{\smallskip}
\hline
\end{tabular}
}
}
\end{table*}


\begin{table*}[ht]
{
\caption{Same as Table 5 for the field galaxies.}
\label{tab_campo}
\centerline{
\begin{tabular}{rccccccc|rcc}
\noalign{\smallskip}
\hline
\noalign{\smallskip}
Field&z&V young&V int.&V old&III young&III int.& I young&D&n$_f$&E(B-V)\\
\noalign{\smallskip}
\hline
\noalign{\smallskip}
121&0.22&  &  & 9&19& 8&64&24 &24&0.1\\
265&0.22&13&44&10& 6&26& 1&10 &34&0.0\\
908&0.34&  &  &  &46&  &54&60 &29&0.0\\
304&0.35&  &  & 8&  &  &92&155&31&0.0\\
203&0.46&27&  &  &  &  &73&24 &21&0.0\\
248&0.46&18&23& 3&  &12&44&27 &38&0.35\\
244&0.47&  &41&19&  &40&  &29 &39&0.0\\
280&0.47&  & 5&31&  &41&23&18 &37&0.0\\
901&0.52&  &  &23&  &  &65&536&28&0.0\\
903&0.53&  & 6&  &  &30&64&34 &32&0.0\\
428&0.54&  &  &49&  & 8&43&88 &28&0.0\\
419&0.55&  &12&  & 5&10&73&22 &31&0.0\\
906&0.59&  &  &14&  &13&73&22 &20&0.2\\
212&0.60&  &  &10&  &51&39&20 &25&0.0\\
286&0.61&  &17& 5&38&17&23&43 &32&0.0\\
400&0.61&  &  &34&  &17&49&183&29&0.0\\
217&0.62&  &  &18&  &38&44&105&42&0.0\\
399&0.68&  &26&22&  & 8&44&157&40&0.0\\
196&0.70&  &19&10&  &40&31& 77&36&0.0\\
346&0.70&  &30& 8&  &51&11&52 &36&0.0\\
199&0.72&  &26&  &  &38&35&76 &35&0.0\\
439&0.80&  &  &35&  & 4&61&183&25&0.0\\
161&0.80&  & 3& 9&  &30&58&117&24&0.0\\
424&0.80&  &  &42&  &30&28&128&11&0.0\\
58 &0.82&  &  &29&  &  &71&82 &23&0.0\\
\noalign{\smallskip}
\hline
\end{tabular}
}
}
\end{table*}


Forty four synthetic spectra were constructed using the solutions given in
Tabs.~\ref{tab_0048} and \ref{tab_campo}. 
Figs.~\ref{espectros_sint} and \ref{espectros_campo_sint} show these spectra
(grey line) superimposed to the observed ones (black line).

\begin{figure*}
\centering
\includegraphics[width=\textwidth]{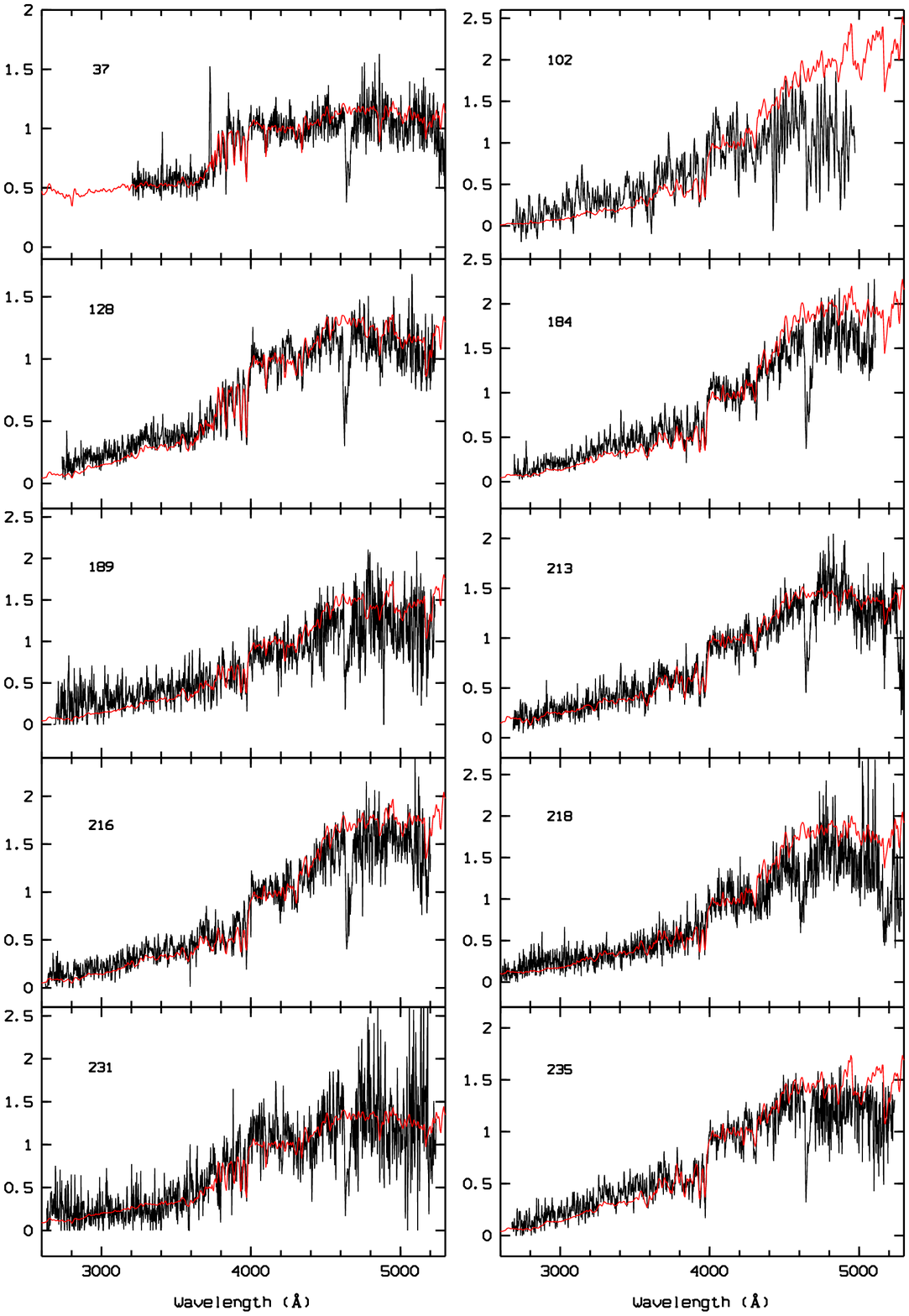}
\caption{Spectra of the cluster galaxies (black line) in the rest-frame
  superimposed to the respective synthetic spectra (grey line)
  computed with the results of the GPG algorithm.
All fluxes are relative values.}
 \label{espectros_sint}
\end{figure*}
\addtocounter{figure}{-1}
\begin{figure*}
\centering
\includegraphics[width=\textwidth]{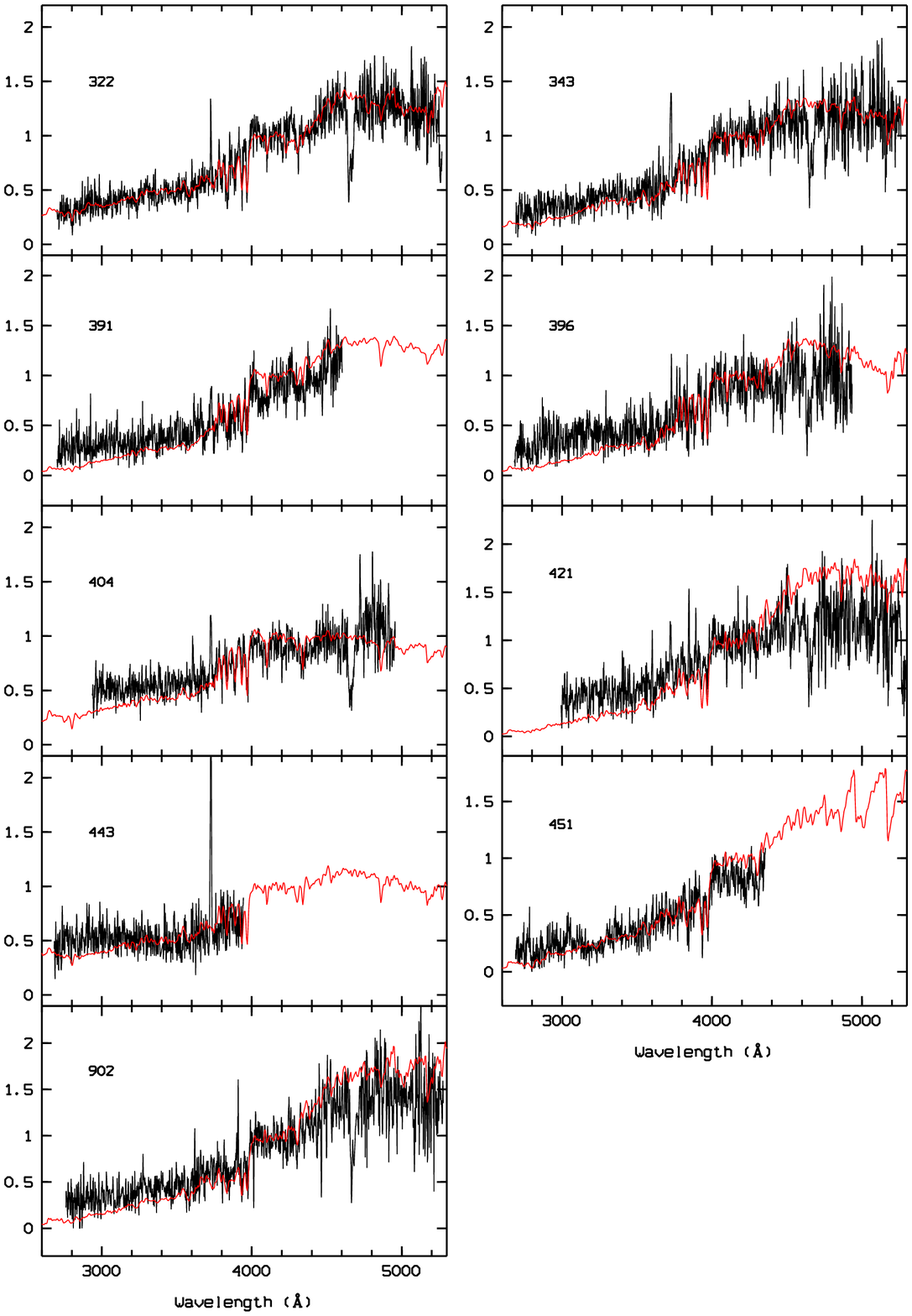}
\caption{(cont.) Rest-frame and synthetic spectra of the cluster galaxies.}
\end{figure*}

\begin{figure*}
\centering
\includegraphics[width=\textwidth]{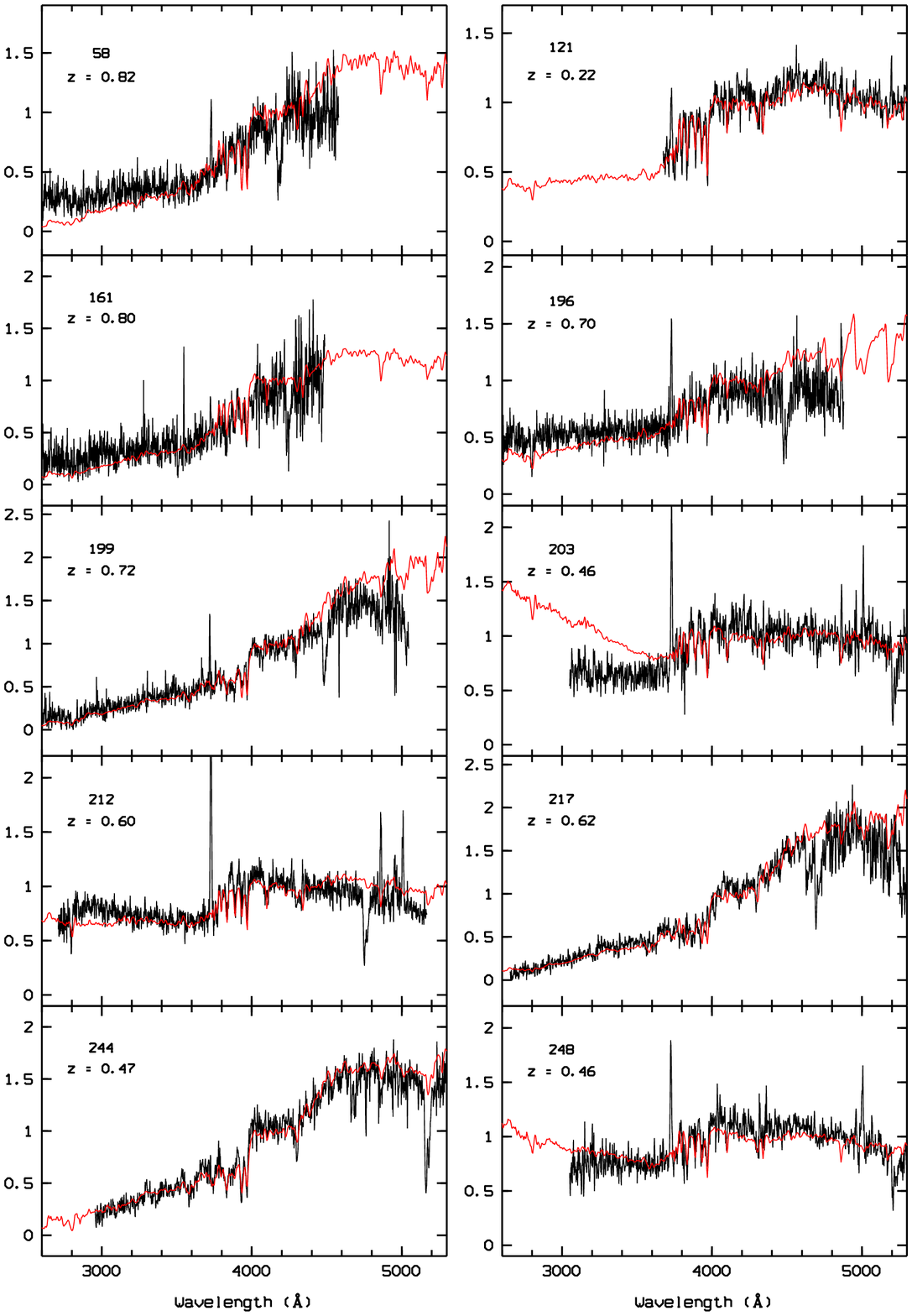}
\caption{Spectra of the field galaxies (black line) in the rest-frame
  superimposed to the synthetic ones (grey line) computed with the
  results of the GPG algorithm. All fluxes are relative values. }
\label{espectros_campo_sint}
\end{figure*}
\addtocounter{figure}{-1}
\begin{figure*}
\centering
\includegraphics[width=\textwidth]{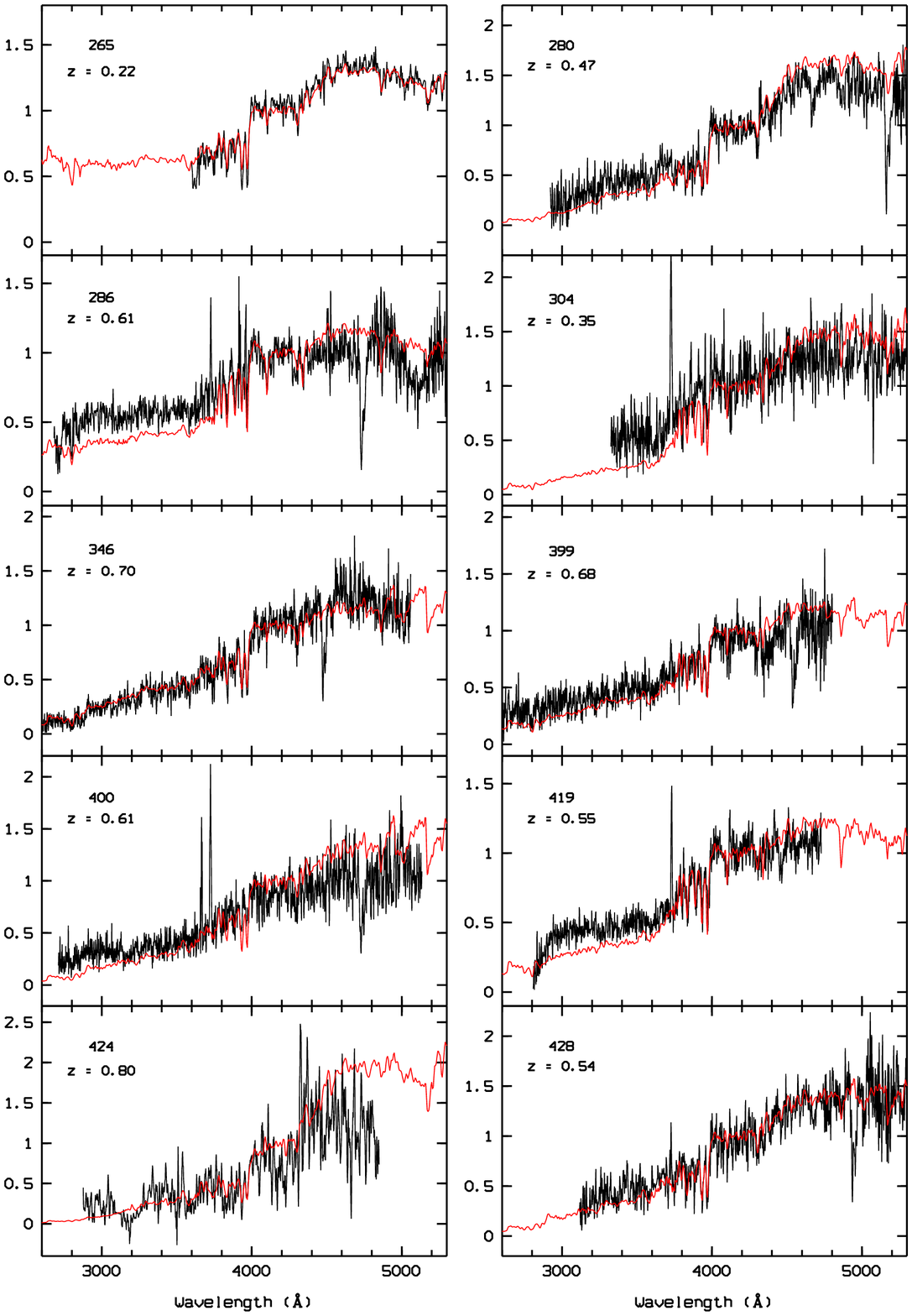}
\caption{(cont.) Rest-frame and synthetic spectra of the field galaxies.} 
\end{figure*}
\addtocounter{figure}{-1}
\begin{figure}
\centering
\includegraphics[width=\textwidth]{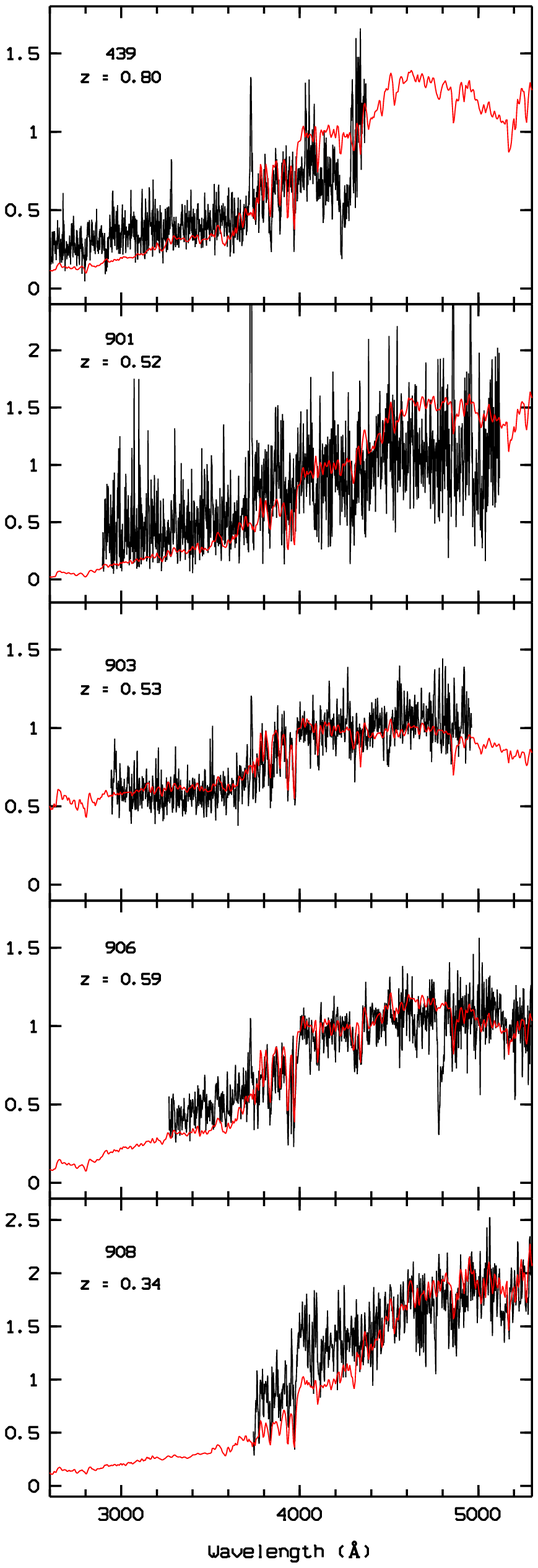}
\caption{(cont.) Rest-frame and synthetic spectra of the field galaxies.}
\end{figure}


Looking more closely at the stellar composition in our sample (see
Tabs.~\ref{tab_0048} and \ref{tab_campo}), we note
that three of the population groups we defined are dominant over the
other three. These are the supergiant component, the intermediate
giant component and the main sequence old one.  We will thus base our
following analysis on these three classes.



\begin{figure}
\centering
\includegraphics[width=9cm]{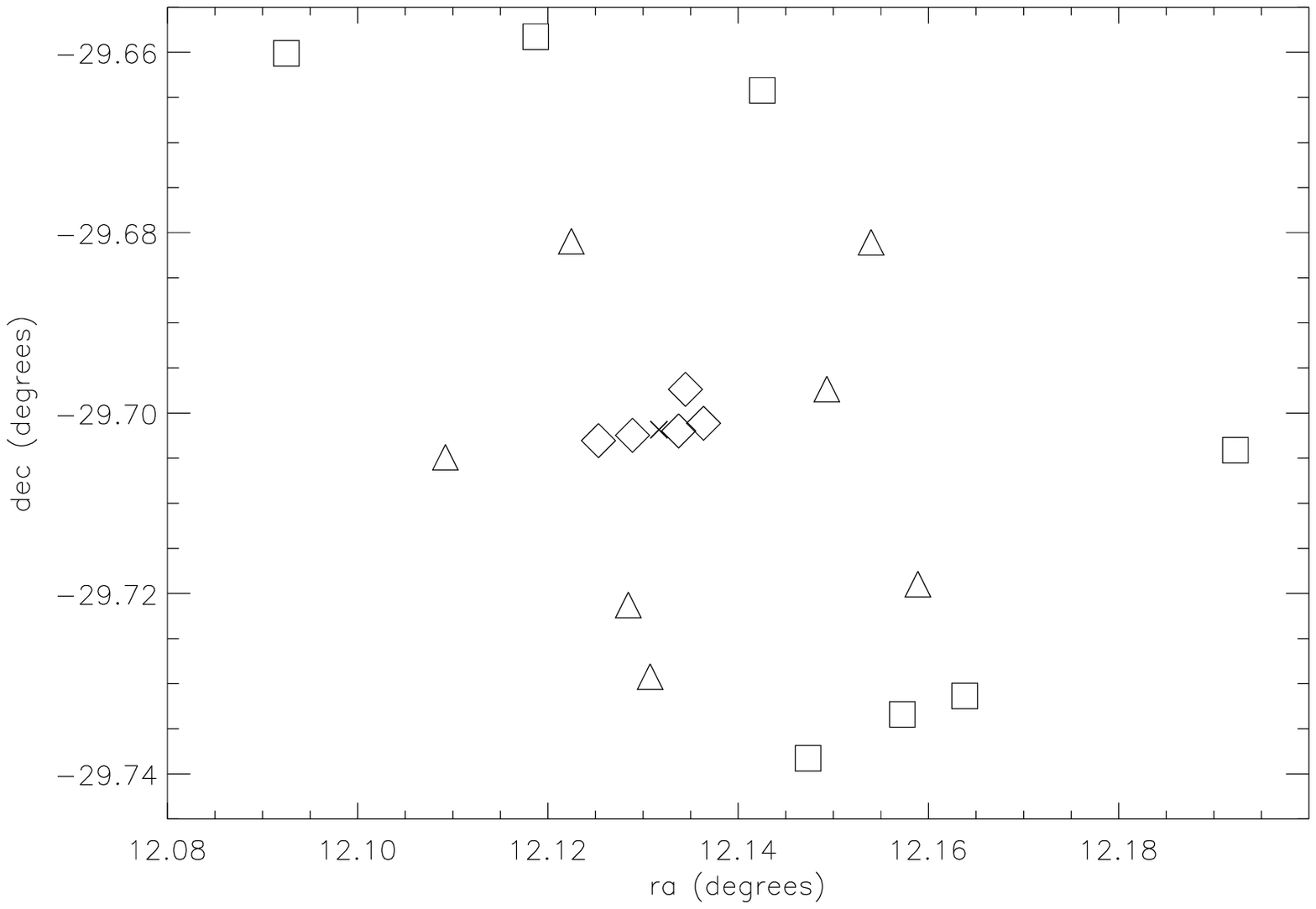}
\caption{Positions of the $19$ cluster member galaxies for 
  which the population 
  synthesis method converged. Each symbol denotes belonging to one of
  the $3$ different cluster regions, set according to projected
  distance to the centre (marked by a cross): from diamonds, through
  triangles up to squares we are moving outwards from the core, through the
  intermediate regions and to the outskirts (as far out as the cluster virial
  radius) of projected cluster regions.}  
\label{psg}
\end{figure}

We began by searching for population gradients within the cluster. To
do so, we divided the cluster into three different regions according
to projected distance to the cluster centre.  The galaxies belonging
to each concentric region are identified by each of the three symbols
in Fig.~\ref{psg}.

\begin{figure}
\centering
\includegraphics[width=9cm]{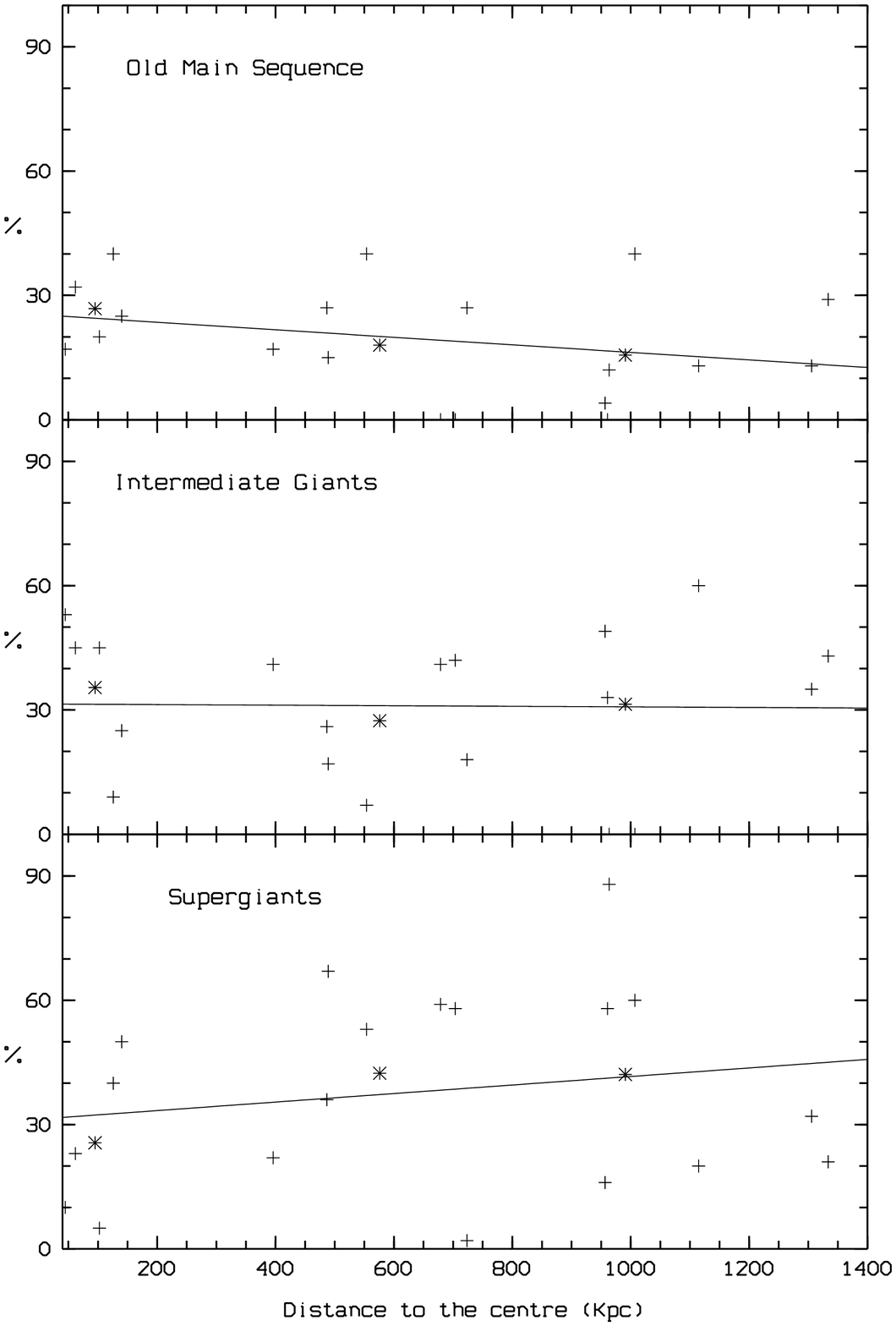}
\caption{Gradient trends found within the cluster (ranging from the 
centre through the intermediate region and far out till the outskirts)
in terms of the three main population components: old main sequence,
intermediate giants and supergiant stars (indicated in
percentages). Asterisks denote the mean of the values of each region.}
\label{media}
\end{figure}

Fig~\ref{media} shows a gradient trend in the stellar population
for two of the three main population classes: old main sequence 
and supergiants.
We have plotted the percentages of these three classes for all galaxies. 
The data points were also polynomial fitted.
Finally, for all galaxies in each group (i.e. centre, intermediate and 
periphery), we have computed the average of their 
population components (shown as asterisks in the figure). 
A population gradient trend is
clearly seen: centre galaxies host mainly intermediate giant stars and
about the same amount of old main sequence and supergiant stars.  As
we progress along the cluster towards the outskirts, the old main
sequence stars become less abundant while the percentage of
supergiants increases. As for the intermediate giant stars no clear
gradient is observed, though is seems that their
number decreases somewhat towards the periphery of the cluster.  We
can conclude that galaxies in the cluster core host older stars
whereas the stellar populations of galaxies inhabiting the outer
cluster regions are dominated rather by young, less evolved stars,
i.e.  supergiants; this means that star formation is predominantly
taking place in the outskirts of the cluster.

\begin{figure}
\centering
\includegraphics[angle=-90,width=9cm]{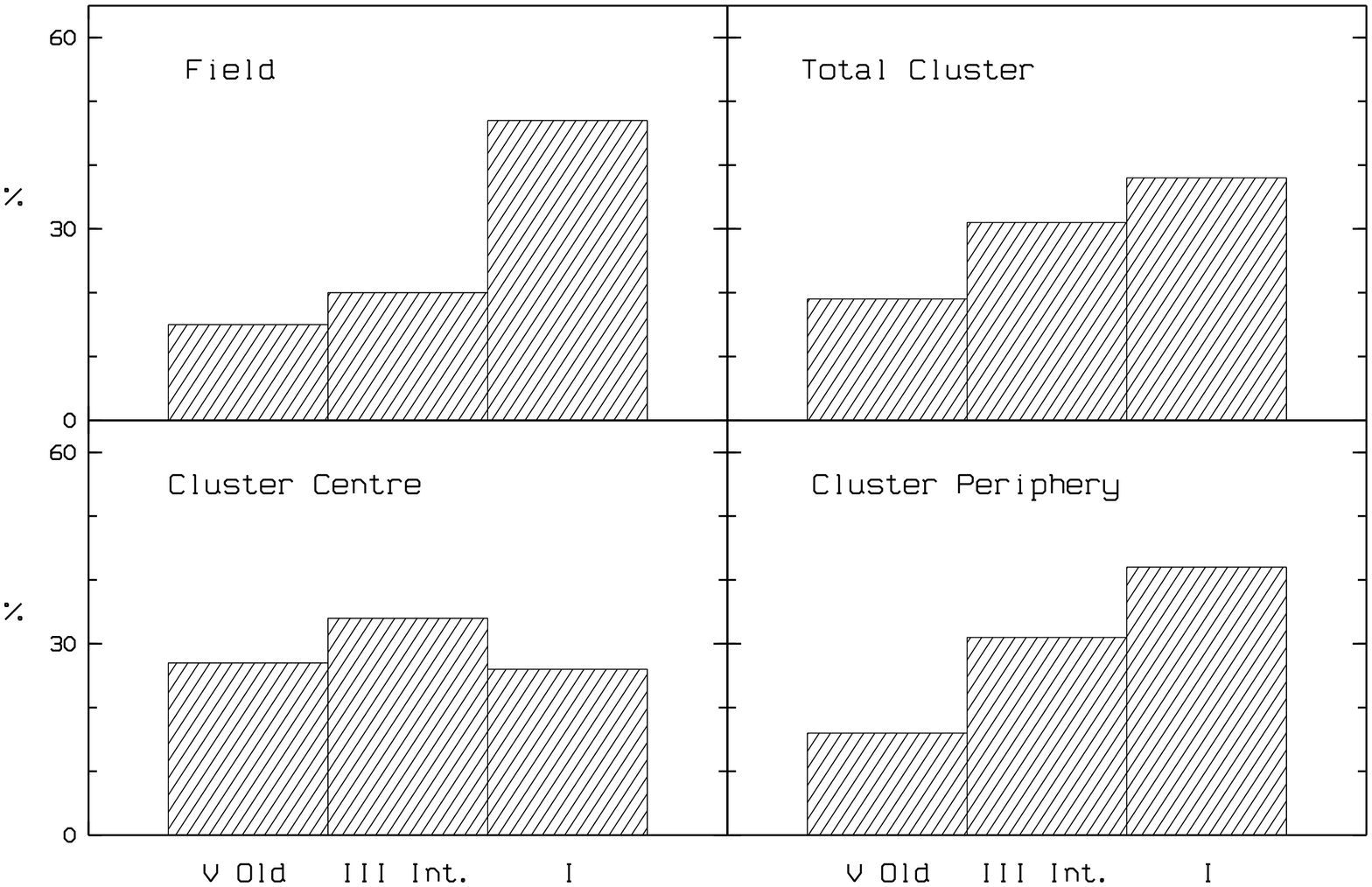}
\caption{Comparison between the population found: in the field galaxies (upper
left); in the cluster as a whole (upper right); in the cluster centre
(bottom left) and in the cluster outskirts (bottom right). The
standard deviations of the values of these averages  vary between 
13\% and 23\%.}
\label{hist_pop}
\end{figure}

The stellar population of the field galaxies seems to be,
in a general way, less evolved that the one found in cluster
members. In fact, in terms of ages, young supergiant stars dominate
the spectra of field galaxies whereas cluster galaxies host a dominant
number of old and intermediate age stars.  Fig.~\ref{hist_pop} shows histograms
comparing the three main population components for the field galaxies
and for the cluster - as a whole, just the centre and finally 
its periphery. We took the averages of the values of the population components
in each region. 
We note that the supergiant population is clearly
decreasing as we move from the field into the cluster and then to its
centre. The old main sequence stars dominate the cluster core with
respect to the outskirts and to field galaxies, whereas the
intermediate giant population seems to be predominantly present in the
cluster (as opposed to what happens in the field).

\section{Analysis on spectral and morphological classes}\label{sec_ka}

\subsection{Currently used methods}\label{normais}

Following the works of Fisher et al. (1998), Balogh et al. (1999), 
Poggianti et al. (1999), Tran et al. (2003) and references therein we
have estimated the percentage of emission line (EL) galaxies and K+A
-type galaxies in the cluster and in the field.

These groups of galaxies were defined on the basis of measures of
particular spectral features: [OII] in emission and the Balmer series,
namely the H$\delta$, H$\gamma$ and H$\beta$ lines when available.

Table \ref{tab_bandas} shows the bandpass used to measure each feature
along with the blue and red -sidebands adopted to determine the local
continuum (as in Fisher et al. 1998). This continuum level was estimated
by fitting a straight line to the flux in the continuum regions.
Equivalent widths were measured using MIDAS.  The error due to the
uncertainty on where one places the continuum is dominant over all other
measurements and statistical uncertainties.  For the absorption
features we measured, this leads to variations on the EW between 
$\pm 1$ and $\pm 3$ \AA\ in
absolute value, for faint and strong lines, respectively.  Whenever
the [OII] emission is significant its EW is often underestimated by as
much as 9 \AA\ due to this uncertainty.
For the weak [OII] lines this variation is always less than
4 \AA\ in absolute value.

\begin{table}[ht]
\caption{Line strength index definitions (from Fisher et al. 1998).}
\begin{tabular}{cccc}
\hline
Feature & Bandpass & Blue sideband & Red sideband \\
\hline
[OII]     & 3716.3--3738.3 & 3696.3--3716.3 & 3738.3--3758.3 \\
H$\delta$ & 4083.5--4122.3 & 4017.0--4057.0 & 4153.0--4193.0 \\
H$\gamma$ & 4319.8--4363.5 & 4242.0--4282.0 & 4404.0--4444.0 \\
H$\beta$  & 4847.9--4876.6 & 4799.0--4839.0 & 4886.0--4926.0 \\
\hline
\end{tabular}
\label{tab_bandas}
\end{table}

Galaxies were classified as emission line (EL), normal (N) and
post-star forming (K+A) galaxies.  Emission line galaxies are those
having an equivalent width (EW) of the [OII] emission line equal to or
higher than 10 $\AA$. K+A are those presenting both EW([OII]) $< 5
\AA$ in emission and Balmer absorption features following at least one
of the next criteria: EW(H$\delta$) $> 5 \AA$ or (EW(H$\delta$) +
EW(H$\gamma$)) $/ 2 > 4 \AA$ or (EW(H$\delta$) + EW(H$\gamma$) +
EW(H$\beta$)) $/ 3 > 4 \AA$. Normal galaxies are all the others that
do not fit into any of these classes.

Table \ref{tab_classes} lists the EWs measured (in \AA) as well as the
criteria used to define the K+A class, spectral classification,
relative position in the cluster and morphological classes given by La
Barbera et al. (2003) based on a luminosity profile analysis, performed on 
VLT/FORS2 images; in what
concerns this last entry, we visually identified our galaxies with the
map they present (in Fig.9 of their paper) though not all of our
galaxies were analysed by La Barbera and collaborators, thus lacking
this classification.

\begin{table*}
\caption{Classifications for all $54$ galaxies having a redshift measure.}
\label{tab_classes}
\centerline{
\begin{tabular}{rccccccccc}
\hline
gal id\#&EW[OII]&EW H$\delta$&EW H$\gamma$&EW H$\beta$&EW (H$\delta$+H$\gamma$)/2 &EW (H$\delta$+H$\gamma$+H$\beta$)/3 & class &  pos. &   morph.\\   
\hline
\underline{\bf Cluster:}&&&&&&&&\\
35   &   23.4 &  8.2  &  1.3&  -9.9&   4.75&    ----&              EL&    p&   --- \\
37   &   12.3 &   4.7 &  4.1&  -6.9&    4.40&      ----  &            EL&    p  &   --- \\
102  &   7.0   &  4.6  &  2.0  &   0.0   &   3.3     &    ----       &         N    &   it  &   sph  \\
128  &   1.2   &  6.0  &  ---   &  3.6   &   ---    &     ----       &         K+A   &  it  &	disk \\
184  &   0.0   &  1.5  &  0.0  &   1.6  &    0.75  &      1.0     &            N   &    it  &   sph \\
189  &  0.2  &   2.1  &  0.0  &   2.2   &   1.05  &      1.4    &             N    &   ct   &   sph \\
213  &   0.0  &   0.3  &  0.5  &   2.4   &   0.4    &     0.9     &            N   &    ct   &   sph \\
216  &    0.0  &   0.0   & 0.2  &   1.1   &   0.1   &      0.4     &            N    &   ct   &   sph \\
218  &   0.0  &   0.0  &  5.7  &   2.7    &  2.8    &     2.8    &             N    &   ct    &  sph \\
231  &    0.0  &   3.4   &  5.5  &   0.0  &    4.45  &      3.0    &             K+A  &   it  &   disk \\
235  &    0.9 &    1.0  &  1.1   &  2.7    &  1.05   &     1.6     &            N     &  ct    &  sph \\
296  &  19.7   &  4.6  &  2.8  &  -4.6   &   3.7    &     ---    &             EL  &    p   &  --- \\
322  &    5.9  &   3.9 &   2.8  &   2.2   &   3.35   &     3.0     &            N   &    it  &   sph \\
343  &  19.0  &   2.1  &  0.5   & -10.3  &   1.3    &     0.9    &             EL   &   it  &   sph \\
391  &    9.4  &   4.9 &   6.9   &  ---   &   5.9   &      ---     &            N   &    it   &  sph \\
396  &    9.1  &   3.7  &  0.0  &   0.0   &   1.8    &     1.2     &           N/K+A    &   p   &  disk \\
404  &  10.0  &   3.9  &  6.2  &   0.0   &   5.0   &      3.4    &             EL  &    p   &  disk \\
421  &    8.7  &   1.8  &  2.7  &   0.0   &   2.2    &     1.5    &             N   &    p &    sph \\
443  &   35.5  &   --- &   ---  &   ---   &   ---   &      ---    &             EL   &   p   &  --- \\
449  &   28.0  &   4.9 &  -1.5  &  -1.7   &   ---    &     ---     &            EL  &    p  &   --- \\
451  &    0.0  &   3.1  &  ---   &  ---  &    ---   &      ---     &            N    &   p  &   --- \\
452  &   34.4  &   1.9 &  -4.2   &  ---    &  ---    &     ---     &            EL   &   p  &   --- \\
902  &    2.8  &   2.2  &  0.9   &  0.0   &   1.5    &     1.0    &             N    &   p   &  sph \\
\hline
\underline{\bf Field:}&&&&&&&&&\\
44   &  --- &   3.8 &  1.3 &  -0.8  &   2.5   &     ---    &            EL    &     &  --- \\
58   &  10.7 &   6.9 &  1.8 &   ---  &   4.4    &    ---     &           EL     &     & --- \\
76   &  32.6 &   2.6 &  3.6  & -6.4  &   3.1    &    ---     &           EL    &      & --- \\
121  &   9.4 &   3.1 &  2.2 &  -0.8  &   2.6   &     ---     &           N      &     & --- \\
161  &   4.9 &   4.2 &  7.0 &   ---   &  5.6   &     ---     &           K+A   &      & --- \\
188  &  69.0 &   0.0 &  4.5 &  -24.0 &   ---   &     ---     &           EL    &      & --- \\
196  &  17.1  &  4.8 &  2.9 &  -13.2 &   3.8   &     ---     &           EL    &      & disk \\
199  &   9.2 &   0.6 &  0.0 &   3.6   &  0.3   &     1.4      &          N     &      & sph \\
203  &  23.0 &   6.0 &  3.7 &  -2.4  &   4.8   &     ---     &           EL    &      & --- \\
212  &  31.5 &   4.9 &  2.0 &  -7.5  &   2.3    &    ---     &           EL    &      & --- \\
217  &   0.0 &   0.0 &  0.1 &   1.1  &   0.05  &     0.4      &          N     &      & sph \\
244  &   2.0 &   0.0 &  1.6  &  3.7 &    0.8   &     1.8     &           N     &      & --- \\
245  &  10.8  &  7.9  & 4.7 &   ---  &   6.3   &     ---     &           EL     &     & --- \\
248  &  17.7 &   1.8 &  0.3 &  -1.1  &   1.0   &     ---     &           EL    &      & --- \\
265  &   0.0 &   2.3 &  2.4 &   2.5  &   2.35  &     2.4     &           N     &      & --- \\
266 &   14.0 &   4.9 &  5.5 &   ---  &   5.2   &     ---     &           EL    &      & --- \\
280  &   0.0  &  1.5 &  0.0 &   3.3  &   0.75  &     1.6    &            N      &     & --- \\
286  &   7.8 &   6.5 &  ---  &  2.6  &   ---   &     ---      &          N     &      & disk \\
304  &  25.4 &   6.9  & 0.0 &  -4.9  &   3.0   &     ---    &            EL    &      & --- \\
346  &   0.0 &   1.6 &  0.9 &   2.8 &    1.2   &     1.8   &             N     &      & sph \\
399  &   5.4 &   --- &  --- &   ---  &   ---   &     ---    &          N/K+A      &       & disk \\
400  &  27.4 &   5.1  & 2.0 &  -0.5  &   3.6   &     ---     &           EL     &     & sph \\
419  &  12.7 &   2.8 &  4.7 &   ---  &   3.8   &     ---     &           EL    &      & sph \\
424  &   0.0 &  -2.7 &  0.0 &   ---  &   ---    &    ---     &           N      &     & sph \\
428  &   3.1 &   0.7 &  4.3 &   1.3  &   2.5    &    2.1    &            N     &      & sph \\
439  &  17.9 &   4.4 &  2.5 &   ---  &   3.4   &     ---     &           EL    &      & disk \\
901  &  53.6 &  10.6 &  0.7 &  -13.0 &   5.6   &     ---     &           EL    &      & sph \\
903  &   7.8  &  3.3 &  3.5 &  -0.2  &   3.4  &      ---     &           N     &      & sph \\
906  &   5.6  &  3.9 &  --- &   ---  &   ---   &     ---     &           N     &      & --- \\
907  &  30.7 &   2.9 &  0.0 &  -9.2  &   1.4   &     ---     &           EL    &      & sph \\
908  &   0.0 &   3.0 &  2.2 &   3.7  &   2.6    &    3.0     &          N/K+A     &      & --- \\
\hline
\end{tabular}
}
Notes: All EW are given in \AA; Column ``class'' gives the spectral 
classification. For three of the objects (\#396, \#399, \#908), since 
the  two methods  described (see sections~\ref{normais} 
and \ref{alternativo}), 
do not give the same result, we display both of them.
Column ``pos.'' indicates the relative position of each 
cluster galaxy within the cluster as in Fig.~\ref{psg} (``ct'' stands for 
centre or core; ``it'' denotes the intermediate region and ``p'' identifies 
peripheral galaxies. Finally, column ``morph.'' lists the morphological 
classification attributed by La Barbera et al. (2003): ``sph'' (for spheroidal 
galaxies) and ``disks''.
\end{table*}

We note that our spectra were not corrected from telluric absorption
so, in some cases, the equivalent widths of the Balmer lines were
contaminated by this effect. Whenever this happened, the respective
line was not considered for our criteria (no entry in 
Table~\ref{tab_classes} is given).

Table~\ref{tab_percentagens} sums up the main results, in terms of
percentages, regarding the fractions of the different spectral-type
galaxies within the cluster and in the field.

\begin{table}
\caption{Number and percentages of the different galaxy classes found
in the cluster (as a whole and in different regions) and in the
field.}
\begin{center}
\begin{tabular}{clll}
\hline
{\bf Cluster}      & EL       & K+A     & N         \\
\hline
all          & 8 (35\%)  & 2 (9\%)  & 13 (56\%)  \\
centre       & 0         & 0        &  5 (100\%) \\
intermediate & 1 (14\%)  & 2 (29\%) &  4 (57\%)  \\
periphery    & 7 (64\%)  & 0        &  4 (36\%)  \\
\hline
{\bf Field}        & 16 (52\%) & 1 (3\%)  & 14 (45\%) \\
\hline
\end{tabular}
\end{center}
\label{tab_percentagens}
\end{table}

With regard to the cluster, one can definitely observe some trends:
we find larger percentages of normal galaxies as we move from the
outskirts to the centre, whereas the number of emission line galaxies
decreases towards the central regions. This agrees with previous
results derived for example from the ENACS survey for lower-z
clusters, where emission line galaxies are found to be more abundant
towards the outskirts of clusters (Biviano et al. 1997).  The K+A
galaxies inhabit the intermediate cluster region.  Such overall trends
have been also identified in the CNOC1 sample of 15 clusters by
Ellingson et al. (2001) using a principal component analysis method.

Looking more closely at the K+A class, and having calculated its
relative frequency and Possionian error in the same way as other
authors, we can compare our results (9\% $\pm$ 6\% of this type)
directly to theirs. Balogh et al. (1999) find an average of 4.4\%
$\pm$ 0.7\% K+A galaxies 
in their sample of 15 clusters at 0.18 $<$ z $<$
0.55, using selection criteria similar to ours, and the same measuring
method (we note that Tran et al. 2003 argue that these fractions
should probably be somewhat higher though).

Dressler et al. (1999) report a mean value of $\sim$ 20\% of k+a and
a+k (i.e. moderate and strong Balmer absorption galaxies without
emission) for a sample of 10 clusters ranging from z=0.37 to z=0.56.
However, their selection criterium differs from the usually adopted one
in the sense that every object with an EW of the H$\delta$ line larger
than 3 \AA\ is considered in this class, probably justifying to some
extent their higher percentages.

For 3 clusters at redshifts 0.33, 0.58 and 0.83, Tran et al. (2003)
determine a fraction of E+A galaxies of 9\% $\pm$ 2\%, 7\% $\pm$ 2\%
and 16\% $\pm$ 3\% respectively, using exactly the same method we
followed here.

It is hard to draw any definite conclusion when comparing our results
to these (and other) works. Even if the method for measuring spectral
features is the same, the magnitude limits, spectroscopic
(in)completeness, quality of the spectra, spatial coverage of the
clusters and possibly other factors are bound to play a role, and the
contribution of such parameters has not yet
been investigated.\\

\subsection{An alternative method for 
identifying post-star forming galaxies}\label{alternativo}

The method we have just used to measure the EW of the lines, though being the
one used by other authors and thus allowing direct comparisons to be
made, does not seem to us as being a reliable one 
as the continua used for the measures are not well defined in
many cases. The EW of the features are normally underestimated, as the
continuum traced is too low (or overestimated in case of a too high
local continuum, which happens in fewer cases).

The criteria that are used intend to find the galaxies with strong
absorption Balmer lines, which appear mainly in A-type stars. In our
work, since we performed stellar population synthesis for almost all
galaxies observed, both in the cluster and in the field, we have
direct access to this information. So, it seems much more natural to
search for these stellar types through the synthesis.  We have thus
viewed several stellar spectra and decided, by means of the relative
strength between the CaII H+K, H$\delta$, H$\gamma$ and the G-band, in
which stellar types the Balmer lines are important features, i.e.
with EW $\geq 6 - 7$ \AA. We
concluded that they appear quite intensely not only in A-type stars
but also for some B and F types: in fact, stars from B5 to F2 spectral
types show important absorption of the Balmer lines.  We have then
searched in the solutions of the synthesis for galaxies presenting 
these types of stars in a large enough percentage 
so that these lines are still strong in the final synthetic spectra
(see Figs.~\ref{espectros_sint} and \ref{espectros_campo_sint}).  We
found that this is the case for percentages equal to or higher than
30\%, so we established this value as our threshold.  Having
identified the galaxies that pass this criterion we then eliminated
those with significant emission in their spectra.

This led us to a final sample of three K+A galaxies in the cluster
and three in the field.  Of these, three had already been found by
applying the criteria of the other authors (\#128 and \#231 within the
cluster, \#161 in the field) and 3 are new ones (\#396 in the cluster
peripheral region, \#399 and \#908 in the field).

We are convinced that the criteria generally used to identify these
galaxies are not good, leading to an underestimate of the absolute
number of existing K+A galaxies. We believe that this is mainly due to
the way of measuring the Balmer lines EWs which, in many cases,
underestimates the line strengths.  Some authors (Dressler et
al. 1999) had already noticed this problem and had pointed to the fact
that in some cases ``strange continua" lead to ``strange measures" of
EW, these appearing in emission (negative values of EW) when in fact
the lines are well seen in absorption.  We subscribe to this view and add
that, due to this problem, K+A galaxies are often lost.

Comparing our results of K+A percentages obtained with the two methods
(i.e. through means of the population synthesis and by the standard
criteria), within the sample of galaxies that
we observed in cluster CL~0048-2942, 13\% are K+A galaxies (contrarily
to the 9\% previously estimated), whereas the field has 10\% of
these galaxies (instead of 3\%).

Table~\ref{tab_percentagens_nova} gives the new classification results
when applying the new method.

\begin{table}
\caption{Same as Table~\ref{tab_percentagens} but using the new 
classification method,
based on the stellar population synthesis results.}
\begin{center}
\begin{tabular}{clll}
\hline
{\bf Cluster}      & EL       & K+A     & N         \\
\hline
all          & 8 (35\%)  & 3 (13\%)  & 12 (52\%)  \\
centre       & 0         & 0        &  5 (100\%) \\
intermediate & 1 (14\%)  & 2 (29\%) &  4 (57\%)  \\
periphery    & 7 (64\%)  & 1 (9\%)    &  3 (27\%)  \\
\hline
{\bf Field}        & 16 (52\%) & 3 (10\%)  & 12 (38\%) \\
\hline
\end{tabular}
\end{center}
\label{tab_percentagens_nova}
\end{table}

Regarding morphology, and going back to Table~\ref{tab_classes}, we
find that all galaxies in the centre (N galaxies) are spheroids; in
the intermediate region, the two K+A galaxies are disks, the EL is a
spheroid, and the remaining ones (N galaxies) are all spheroids. As
for the periphery, morphological classification is available for four
galaxies only: two are spheroids (N galaxies) and two are disks (one
EL and one K+A). So, in terms of a morphological gradient, spheroids
are more abundant in the centre and their number decreases as one
moves to the outskirts of the cluster, being replaced by disk
galaxies.  Quantitatively, we find 100\% of spheroids in the centre,
71\% of spheroids and 29\% of disk galaxies in the intermediate region,
while the outer region presents 50\% of disk galaxies and 50\% of
spheroids even though this last result must be taken with caution
since we lack the morphological classifications for six of the
galaxies. However, La Barbera et al. (2003), with a different cluster
sampling, also find a greater central concentration of spheroids
whereas disks are predominantly located towards the
outskirts of the cluster, in regions of lower galactic density.\\

As for field, and contrarily to what happens in the cluster, N
galaxies are much less frequent, being outnumbered by EL ones. This
kind of result has been observed long ago (see e.g. Balogh et al.
1997 and references therein) and remains evident in the latest surveys
(e.g. Goto et al. 2003). 

For the EL field galaxies, their EW([OII]) reaches higher values (up
to twice as much) than in their cluster counterparts leading us to
believe that star formation is more intense in the field - or, rather,
suppressed within the cluster. Besides, and as far as we have probed
the cluster periphery (where the probability of observing infall
galaxies is higher), no enhancement of the star formation is detected
in cluster members.  Evidence for lower star formation rates within
the virial radius for intermediate to high -redshift clusters
relatively to the coeval field had already been reported by e.g. Balogh
et al. (1997, 1998, 1999), Ellingson et al. (2001), Postman et al.
(2001), Kodama \& Bower (2001), Lewis et al. (2002), G\'omez et al.
(2003). Regarding our work, though, two caveats need to be
taken into account before drawing any conclusions: (1) we do have
limited spectroscopic coverage, which may have caused us to miss a
fraction of star forming galaxies - farthest away from the cluster
core - that equal or even supersede the coeval field counterparts (in
number and/or in star forming activity); (2) our field galaxy sample
is not coeval with the cluster galaxies, hampering direct
comparisons. However, we note that it is equally composed of
foreground and background galaxies: $15$ and $16$, respectively; in
each of these two groups, $6$ and $11$ galaxies are ELs and the
average EWs of the [OII] emission are $16.8$ \AA\ and $30.2$ \AA,
respectively.  Instead, the $8$ cluster ELs produce an average EW of
the [OII] emission line of $22.8$ \AA.  The only thing one can say
from these figures is that for those galaxies within the virial radius
of CL~0048-2942 undergoing star formation, we would expect their
activity to be equal to or lower than the coeval field galaxies.
Whether the fraction of EL galaxies is higher or lower than in the
coeval field cannot be ascertained from our observations.

Furthermore, it is interesting to note that all the K+A galaxies we
identified in our samples (cluster and field) having a morphological
classification by La Barbera et al. (2003) (4 in a total of 6) are
disk galaxies. This agrees with previous findings that the majority of
these systems seem to be disks (Kelson et al. 200, Tran et al. 2003
and references within these papers).\\

One last consideration concerning the results presented in this
section and in the previous one: we are aware that projection effects
may have an influence on our results, but our limited amount of data
on cluster members prevents us from making any correction; the
conclusions we reached should thus be taken rather as indications
only. However, we recall that de-projecting distances is heavily
dependent on the cluster density profile one assumes - a source of
uncertainties that are hard to quantify -, and is expected not to
change results qualitatively (Ellingson et al. 2001).  Still, a larger
amount of redshifts would enable us to perform a complementary
line-of-sight approach, thus giving us a full and desirable
three-dimensional view of the galaxy population properties of
CL~0048-2942.

\section{Colours}\label{sec_conc}

\begin{figure}
\centering
\includegraphics[width=9cm]{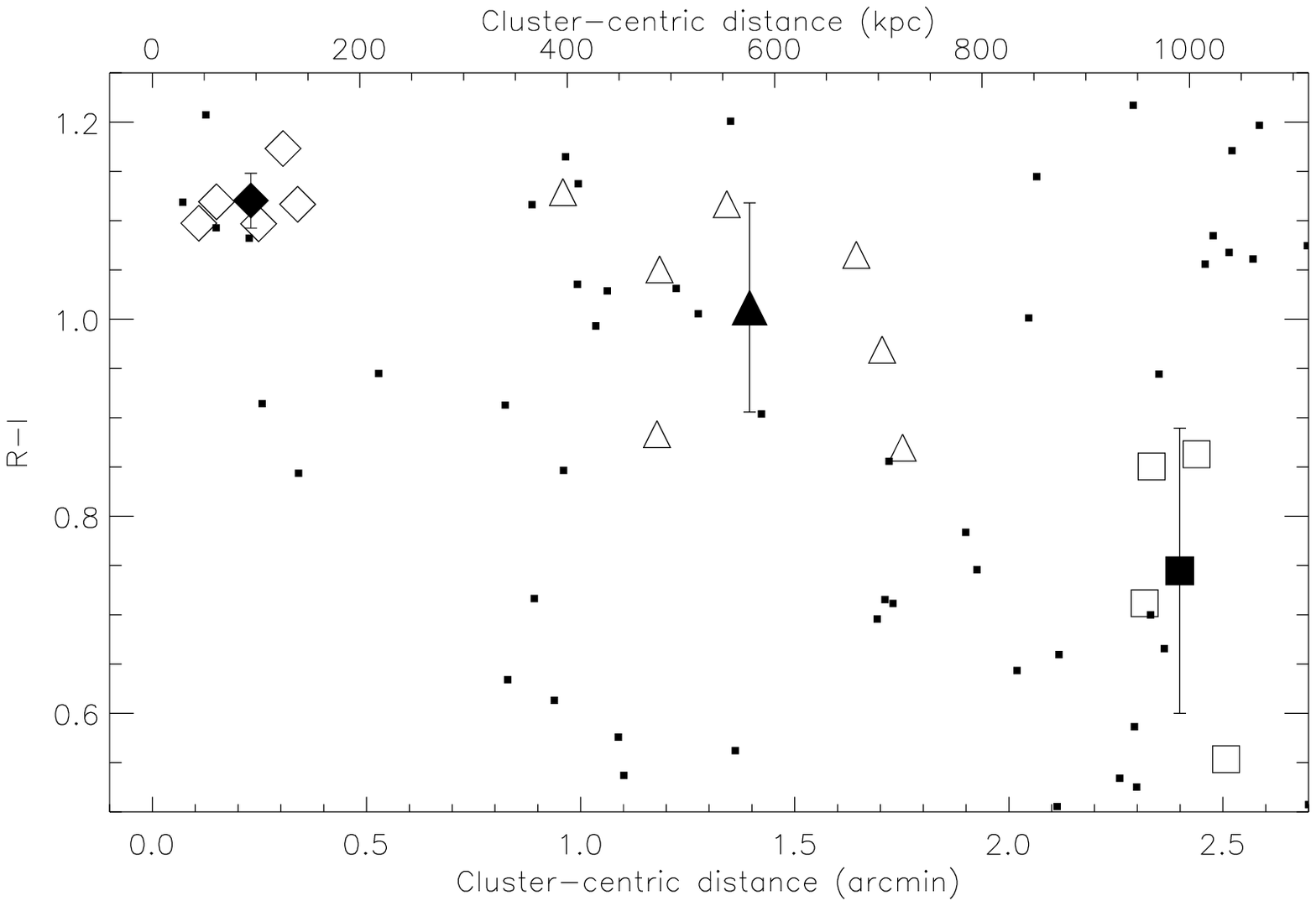}
\caption{R-I colour as a function of cluster-centric projected distance 
  for all cluster galaxies having reliable photometry. For these objects
the mean value of the error in the R-I colour is $\sim$ 0.03 mag. 
Open symbols
  are the same as in Fig.~\ref{psg}, indicating the cluster-centric
  distance class in which a galaxy is considered. Filled symbols give,
  at mean distance for each class, the mean colour of that class (the
  ``error bar'' indicates the dispersion of the values used to compute
  the average). Smaller dots refer to field galaxies.}
\label{colors}
\end{figure}

\begin{figure}
\centering
\includegraphics[width=9cm]{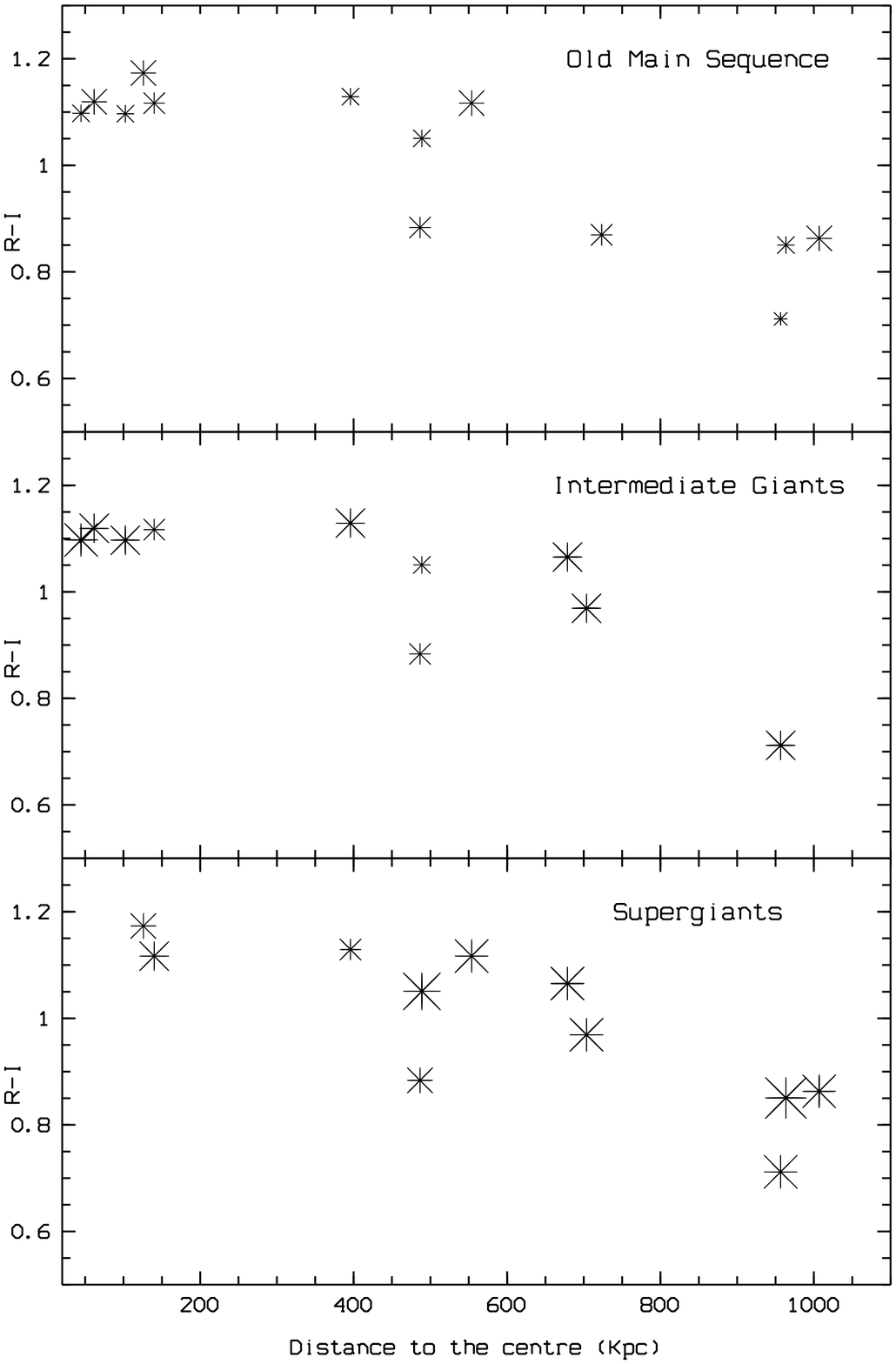}
\caption{Colours versus distance to the 
centre of the cluster. The symbols sizes are proportional to the
percentages of the 3 stellar population components indicated in each panel.}
\label{pop_colors}
\end{figure}

The galaxy population gradient revealed by our stellar population
synthesis analysis (and presented in section~\ref{sec_res}) is
confirmed by the colours of the cluster members, once we distribute
them into the same classes as before (centre, intermediate and
periphery), which can be translated into cluster-centric projected
distances (see Fig.~\ref{colors}).  Central galaxies have in average
the colour of the red sequence, R$-$I$ = 1.12$ magnitudes, that
signals the cluster presence in a colour-magnitude plot, as determined
by Andreon et al. (2004).  Then, as one moves away from the centre of
the cluster, there is a clear trend of blueing of the stellar
population. The change of galaxy colour as a function of radius has
been observed since the works of Butcher \& Oemler (1978, 1984) within
the same physical region we are probing in CL~0048-2942 and even till
further out (Kodama \& Bower 2001; Pimbblet et al. 2002).
We have also plotted the stellar population parameters in terms of the 3
main components (old main sequence, intermediate giants and supergiant
stars) versus colour and distance to the cluster centre 
(see Figure~\ref{pop_colors}). Note the trends in the stellar populations
as we move from the centre towards the outskirts of the cluster.
Regarding colour and stellar populations one can see, by inspection of 
Figure~\ref{pop_colors}, that in fact, supergiants are the stellar class
more abundant in the bluer colours whereas giants and main sequence
stars together are more important for redder ones.

CL~0048-2942 has a fraction of blue galaxies f$_b = 0.29 \pm 0.05$
(Andreon et al. 2004; though see La Barbera et al. 2003 for a
different estimate). Such a value is compatible with the extrapolation
to $z \sim 0.6$ of the Butcher \& Oemler (1984) results, and so could
be providing an indication of a significant content of blue galaxies.
However, a rigorous analysis on the computation of error bars
presented also in Andreon et al. (2004), does not allow to rule out a
constant blue fraction with redshift for the whole Butcher \& Oemler
sample including CL~0048-2942. Apart from demonstrating the
uncertainties that still prevail in establishing the Butcher-Oemler
effect as a fact, those results show that the population of
CL~0048-2942 seems to be just as blue as one would expect from the age
of the galaxies at $z \sim 0.64$.

\section{Summary}\label{sec_sum}

We studied a field centered on cluster CL0048-2942, measuring
redshifts for 54 objects, 23 of them belonging to the cluster with
redshift of around 0.64. The line-of-sight velocity dispersion of
CL0048-2942, based
on this sample, is $680$ $\pm$ 140 km/s.

Our analysis have unveiled some interesting
radial trends in the galaxy population, namely the
existence of (1) a stellar populations gradient, (2) a spectral
classes gradient and (3) a colour gradient.

Our stellar populations synthesis method revealed that centre cluster
galaxies host mainly intermediate and old stars; such a population
gradually seems to change towards the outskirts of the cluster (as far
out as its virial radius, approximately), where supergiants dominate
the spectra of the galaxies, denoting a higher rate of star formation
in this region.  This trend is followed by the colours of the
galaxies: redder in the cluster inner regions and bluer at larger
radial distances.

Field galaxies, analysed with the same method, are found to be predominantly
made of less evolved stars, namely supergiants, with smaller components of
main sequence and giant stars.

The stellar populations, both for cluster and for field galaxies, are
mainly poor in dust content.

In what concerns spectral classes, we used a new method, based on 
the stellar population synthesis results,
that takes into account all possible absorption features in the spectrum
and thus makes optimal use of the data.
We thus find that emission-line galaxies are present
mainly in the field and in the outskirts of the cluster whereas normal
galaxies are concentrated in the centre of the cluster. Concerning K+A
galaxies, 3 were found in the cluster (in the intermediate and peripheral
regions) and 3 in the field.
Comparisons were attempted with the estimates provided by several
authors on the K+A content of different redshift clusters. However,
poor statistics and the unknown contribution of numerous involved
parameters, such as incompleteness, magnitude limit, quality of the
spectra, etc., do not permit to draw any definite conclusions.

We correlated our cluster data with morphological information produced
by La Barbera et al. (2003), concluding that 100
cluster galaxies are spheroids, while there are 71
disk galaxies in the intermediate region. The outer region presents
50
of the galaxies in this region.

To conclude, we believe the trends reported in this paper to be real:
we have no reason to expect our spectroscopic sampling to suffer from
any selection bias, so the results that we have obtained here should
provide a good indication of the whole picture. The radial gradients
that we observe could then be tracing a ``radial evolution'' of the
cluster members and tempt us into inferring some implications on
environmentally-driven changes, probably in the sequence of infall
episodes. On the other hand, it could simply reflect the
morphology-density relation (Dressler 1980).  Choosing between the two
scenarios for CL~0048-2942 would require morphological classification
for all our sampled galaxies as well as for a field sample observed in
the same redshift range, and a larger spatial coverage with complete
spectroscopy.\\

A growing feeling, however, is that local density does seem to play
a key role in spectral characteristics that reveal spectral class
transformations, which can be accompanied - though in different,
longer, time scales - by morphological ones.  The limiting surface
density that marks the separation between field-type and cluster-like
galaxies, at least in terms of star formation, should be around 1
Mpc$^{-2}$, which generally is attained around the virial radius (see
Bower \& Balogh 2004 and references therein), our limit of spatial
coverage for CL~0048-2942.

It would be interesting to push our analyses further, complementing
our results with local density measures (as performed by e.g. Kodama et
al. 2001, Lewis et al. 2002, Pimbblet et al. 2002, G\'omez et al. 2003,
Balogh et al. 2004, Kauffmann et al. 2004) and trying to understand
whether these accompany the gradients we detected. Given the low
numbers we are dealing with in this paper, it wouldn't be sensible to
try and measure a density so as to compare with the works referenced
above.  This work is thus a future prospect for CL~0048-2942.

\begin{acknowledgements}
  We are greatly indebted to Didier Pelat for the use of the
  population synthesis GPG program. We thank Jo\~ao Fernandes for
  fruitful discussions on stellar evolution and for help on computing
  the stellar ages used in this work. It is also a pleasure to thank
  Daniel Folha for his help in taming IDL.
  We would also like to address special thanks to the anonymous referee who did
  an excelent work in enthusiastically revising this paper, making many 
  comments and suggestions that led us to improve this paper.  
  M. Serote Roos acknowledges
  financial support from FCT, Portugal, under grant no. BPD/5684/01.
  I. M\'arquez acknowledges financial support from the Junta de
  Andaluc\'{\i}a and DGIyT grants AYA2001-2089 and AYA2003-00128.  The
  authors acknowledge financial support from Portugal FCT/ESO Project
  ref. PESO/C/PRO/15130/1999.
\end{acknowledgements}


\begin{thebibliography}{}

\bibitem{}Abell G.O. 1958, ApJS 3, 211
  
\bibitem{}Andreon S., Lobo C., Iovino A. 2004, MNRAS 349, 889

\bibitem{}Arnaud M., Evrard A.E. 1999, MNRAS 305, 631

\bibitem{} Balogh M.L., Morris S.L., Yee H.K.C., Carlberg R.G.,
  Ellingson E. 1997, ApJ 488, L75
  
\bibitem{} Balogh M.L., Schade D., Morris S.L. et al. 1998, ApJ 504,
  L75

\bibitem{} Balogh M.L., Morris S.L., Yee H.K.C., Carlberg R.G.,
  Ellingson E. 1999, ApJ 527, 54

\bibitem{}Balogh M., Bower R.G., Smail I. et al. 2002, MNRAS 337, 256

\bibitem{}Balogh M., Eke V., Miller C. et al. 2004, MNRAS 348, 1355

\bibitem{} Blanton E.L., Gregg M.D., Helfand D.J. et al. 2003, AJ 125,
  1635

\bibitem{}Benoist C., da Costa L., J{\o}rgensen H.E. 2002, A\&A 394, 1

\bibitem{}Bertin E., Arnouts S., 1996, A\&AS 117, 393
  
\bibitem{} Best P.N., Lehnert M.D., Miley G.K., R\"ottgering H.J.A.
  2003, MNRAS 343, 1

\bibitem{}Biviano A., Katgert P., Mazure A. et al. 1997, A\&A 321, 84
  
\bibitem{} Bower R.G., Balogh M.L. 2004, in ``Clusters of Galaxies:
  Probes of Cosmological Structure and Galaxy Evolution'', eds. J.S.
  Mulchaey, A. Dressler, and A. Oemler, p. 326

\bibitem{}Bressan A., Fagotto F., Bertelli G., Chiosi C. 1993, A\&ASS 100, 647 

\bibitem{}Brown M.J.I., Webster R.L., Boyle B.J. 2001, AJ 121, 2381

\bibitem{}Butcher H., Oemler A. 1978, ApJ 219, 18

\bibitem{}Butcher H. \& Oemler A. 1984, ApJ 285, 426

\bibitem{}Cardelli J.A., Clayton G.C., Mathis J.S. 1989, ApJ 345, 245
  
\bibitem{}Condon J.J., Cotton W.D., Greisen E.W. et al. 1998, AJ 115, 1693
  
\bibitem{} Couch W.J., Sharples, R. M. 1987, MNRAS 229, 423

  
\bibitem{} Della Ceca R., Scaramella R., Gioia I.M. et al. 2000, A\&A,
  353, 498

\bibitem{} Deltorn J.-M., Le F\`evre O., Crampton D., Dickinson M.
  1997 ApJ 483, L21


\bibitem{}Dressler A., 1980, ApJ 236, 351

\bibitem{} Dressler A., Gunn J.E. 1983, ApJ 270, 7

\bibitem{} Dressler A., Smail I., Poggianti B.M. 1999, ApJS 122, 51

\bibitem{} Dressler A. 2004, in ``Clusters of Galaxies: Probes of
  Cosmological Structure and Galaxy Evolution'', eds. J.S. Mulchaey,
  A. Dressler, and A. Oemler, p. 207
  
\bibitem{} Donahue M., Voit G.M., Gioia, I. et al. 1998, ApJ, 502, 550


\bibitem{}Eales S.A. 1985, MNRAS 214, 27
  
\bibitem{} Ebeling H., Jones L.R., Perlman E. et al. 2000, ApJ 534, 133

\bibitem{} Ebeling H., Jones L.R., Fairley B.W. et al. 2001, ApJ 548, L23

\bibitem{}Ellingson E., Lin H., Yee H.K.C., Carlberg R.G. 2001, ApJ
  547, 609
  
\bibitem{}Fisher D., Fabricant D., Franx M., van Dokkum P. 1998, ApJ 498, 195

\bibitem{}G\'omez P.L., Nichol R.C., Miller C.J. et al. 2003, ApJ 584,
  210
  
\bibitem{} Goto T., Nichol R.C., Okamura S. et al. 2003, PASJ 55, 771

\bibitem{} Goto T., Yagi M., Tanaka M., Okamura S. 2004, MNRAS 348,
  515


\bibitem{} Hartwick F.D.A. 2004, ApJ 603, 108

\bibitem{}Howarth I.D., 1983, MNRAS 203, 301

\bibitem{}Katgert P., Mazure A., Perea J. et al. 1996, A\&A 310, 8

\bibitem{} Kauffmann G. 1995, MNRAS 274, 161

\bibitem{} Kauffmann G., White S.D.M., Heckman T.M. et al. 2004, MNRAS
  submitted or astro-ph/0402030

\bibitem{} Kelson D.D., Illingworth G.D., van Dokkum P.G., Franx M. 2000, 
ApJ 531, 184

\bibitem{} Kennicutt R.C. 1983, AJ 88, 483

\bibitem{} Kodama T., Bower R.G. 2001, MNRAS 321, 18

\bibitem{} Kodama T., Smail I., Nakata F., Okamura S, Bower R.G. 2001,
  ApJ 562, L9

\bibitem{}La Barbera F., Merluzzi P., Iovino A. et al. 2003, A\&A 399, 899

\bibitem{}Landolt A.U. 1992, AJ 104, 340
  
\bibitem{}Lewis I., Balogh M., De Propris R. et al. 2002, MNRAS 334,
  673

\bibitem{} Lilly S.J., Le F\`evre O., Hammer F., Crampton D. 1996, ApJ
  460, L1

\bibitem{} Lima Neto G.B., Capelato H.V., Sodr\'e L. Jr, Proust D.
2003, A\&A 398, 31

\bibitem{} Lobo C., Iovino A., Lazzati D., Chincarini G., 2000, A\&A 360, 896

\bibitem{} Lubin L.M., Postman M.; Oke J.B. et al. 1998, AJ 116, 584

\bibitem{} Lubin L.M., Oke J.B., Postman M. 2002, AJ 124, 1905

\bibitem{} Lubin L.M., Mulchaey J.S., Postman M. 2004, ApJ 601, L9

\bibitem{} Luppino G.A., Gioia I.M. 1995, ApJ, 445, L77  

\bibitem{} Madau P., Ferguson H.C., Dickinson M.E. et al. 1996, MNRAS
  283, 1388

\bibitem{} Madau P., Pozzetti L., Dickinson M. 1998, ApJ 498, 106


\bibitem{}Nonino M., Bertin E., da Costa L. et al. 1999, A\&AS 137, 51 

\bibitem{}Oemler A.,Jr. 1974 ApJ 194, 10

\bibitem{}Pelat D., 1997, MNRAS 284, 365

\bibitem{}Pickles A.J., 1998, PASP 110, 863

\bibitem{}Pimbblet K.A., Smail I., Kodama T. et al. 2002, MNRAS 331, 333

\bibitem{}Poggianti B.M., Smail I., Dressler A. et al. 1999, ApJ 518, 576
  
\bibitem{} Postman M., Lubin L.M., Oke J.B. 1998, AJ 116, 560

\bibitem{} Postman M., Lubin L.M., Oke J.B. 2001, AJ 122, 1125

\bibitem{}Quintero A.D., Hogg D.W., Blanton M.R. et al. 2004, 
ApJ 602, 190

\bibitem{}Ramella M., Biviano A., Boschin W. et al. 2000, A\&A 360, 861

\bibitem{} Rosati P., Stanford S.A., Eisenhardt P.R. et al. 1999, AJ
  118, 76

\bibitem{}Schlegel D.J., Finkbeiner D.P., \& Davis M. 1998, ApJ, 500, 525  


\bibitem{} Smith G.P., Treu T., Ellis R.S., Moran S.M., Dressler A.
  2004, ApJ submitted or astro-ph/0403455

  
\bibitem{}Tonry J., Davis M., 1979, AJ 84, 1511
  
\bibitem{} Tran K.-V.H., Franx M., Illingworth G. et al. 2003, ApJ
  599, 865
  
\bibitem{} Tran K.-V.H., Franx M., Illingworth G. et al. 2004, ApJ
  accepted or astro-ph/0403484

\bibitem{}Valtchanov I., Pierre M., Willis J. et al., 2003, A\&A
  submitted or astro-ph/0305192
  
\bibitem{} van Dokkum P.G., Franx M., Kelson D.D., Illingworth G.D.
  1998, ApJ 504, L17

\bibitem{} van Dokkum P.G., Franx M., Fabricant D., Kelson D.D.,
  Illingworth G.D. 1999, ApJ 520, L95

\bibitem{}Wittman D., Margoniner V.E., Tyson J.A. et al. 2003, ApJ
  597, 218

\bibitem{}Wu X.-P., Fang L.-Z., Xu W. 1998, A\&A 338, 813

\bibitem{}Wu X.-P., Xue Y.J., Fang L.-Z., 1999, ApJ 524, 22

\bibitem{}Yates M.G., Miller L., Peacock J.A. 1989, MNRAS 240, 129

\bibitem{} Yee H.K.C., Ellingson E., Carlberg R.G. 1996, ApJS 102, 269

\bibitem{}Yee H.K.C., L\'opez-Cruz O. 1999, AJ 117, 1985

\bibitem{}Yee H.K.C., Ellingson E. 2003, ApJ 585, 215

\end{thebibliography}
\end{document}